\documentclass[submitorcamera]{vgtc}              

\onlineid{2953}

\usepackage{microtype}                 
\PassOptionsToPackage{warn}{textcomp}
\usepackage{textcomp}
\usepackage{mathptmx}                  
\usepackage{times}                     
         
\usepackage{cite}                      
\usepackage{tabu}                      
\usepackage{booktabs}                  
\usepackage{amsmath}
\usepackage{algorithm2e}
\usepackage{adjustbox}
\usepackage{url}
\usepackage{colortbl}
\usepackage{amsfonts}

\usepackage{float}

\graphicspath{{./pictures/}}

\newlength\mylen

\vgtccategory{Research}

\vgtcpapertype{Evaluation \& Empirical Research Papers}

\title{Interactive Visualization of Time-Varying Flow
Fields\\ Using Particle Tracing Neural Networks}

\author{Mengjiao Han\thanks{e-mail: \{mengjiao,jixianli,beiwang,crj\}@sci.utah.edu}, Jixian Li\footnotemark[1], Sudhanshu Sane\thanks{ssane@sci.utah.edu}, Shubham Gupta\thanks{shubhamg2404@gmail.com}, Bei Wang\footnotemark[1], Steve Petruzza\thanks{steve.petruzza@usu.edu}, and Chris R. Johnson\footnotemark[1]}

\abstract{
\vspace{-0.5em}
Lagrangian representations of flow fields have gained prominence for enabling fast, accurate analysis and exploration of time-varying flow behaviors. 
In this paper, we present a comprehensive evaluation to establish a robust and efficient framework for Lagrangian-based particle tracing using deep neural networks (DNNs).
Han et al. (2021) first proposed a DNN-based approach to learn Lagrangian representations and demonstrated accurate particle tracing for an analytic 2D flow field. 
In this paper, we extend and build upon this prior work in significant ways. 
First, we evaluate the performance of DNN models to accurately trace particles in various settings, including 2D and 3D time-varying flow fields, flow fields from multiple applications, flow fields with varying complexity, as well as structured and unstructured input data. 
Second, we conduct an empirical study to inform best practices with respect to particle tracing model architectures, activation functions, and training data structures. 
Third, we conduct a comparative evaluation of prior techniques that employ flow maps as input for exploratory flow visualization. 
Specifically, we compare our extended model against its predecessor by Han et al. (2021), as well as the conventional approach that uses triangulation and Barycentric coordinate interpolation. 
Finally, we consider the integration and adaptation of our particle tracing model with different viewers. 
We provide an interactive web-based visualization interface by leveraging the efficiencies of our framework, and perform high-fidelity interactive visualization by integrating it with an OSPRay-based viewer. 
Overall, our experiments demonstrate that using a trained DNN model to predict new particle trajectories requires a low memory footprint and results in rapid inference. 
Following best practices for large 3D datasets, our deep learning approach using GPUs for inference is shown to require approximately 46 times less memory while being more than 400 times faster than the conventional methods.}

\keywords{Flow visualization, Lagrangian-based particle tracing, deep learning, neural networks, scientific machine learning\vspace{-0.5em}}

\teaser{
\vspace{-1em}
 \centering
 \includegraphics[width=0.9\linewidth, keepaspectratio]{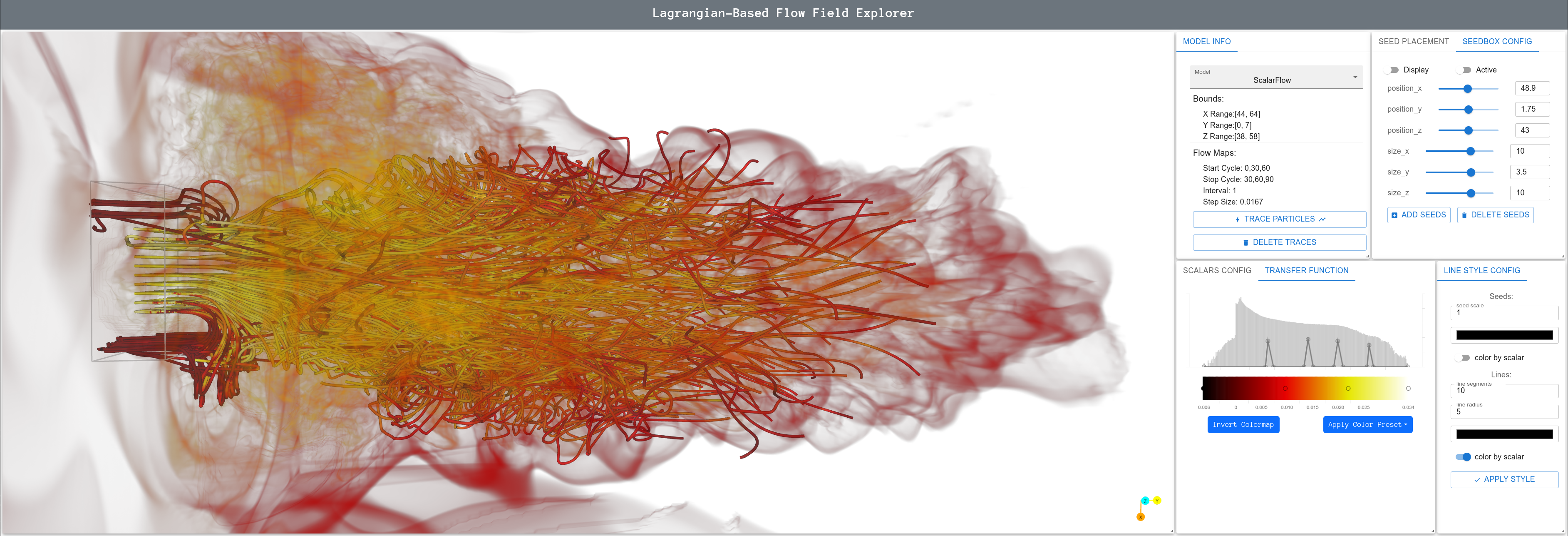}
 \caption{Our web-based visualization interface, integrated with our particle tracing neural networks, enables users to visualize and explore large 3D time-varying flow fields interactively. The interface offers 3D visualization of pathlines and user-uploaded scalar fields, along with a variety of configuration options for seed placement and rendering parameters. In this example, the model trained on the \textbf{ScalarFlow} dataset was used to display pathlines, with the FTLE of the dataset as the background volume. 
 The training dataset was generated using 100,000 seeds placed in the injection area of $[44, 64] \times [0, 7] \times [38, 58]$ over 90 time steps. Three deep learning models, each trained for 30 time steps, were utilized in this case. It took one second to load the models and 2.7 seconds to infer 300 pathlines displayed in the visualization using a dual-socket workstation equipped with an NVIDIA Titan RTX GPU. The trained model's total storage size is 78.6 MB, further reducing the space consumption required for saving the original flow maps by 46-fold.\vspace{-1em}
 }
 \label{fig:teaser}
}

\graphicspath{{figs/}{figures/}{pictures/}{images/}{./}} 

\begin{document}

\firstsection{Introduction}

\maketitle

Time-varying flow visualization is useful for validating, exploring, and gaining insight from computational fluid dynamics simulations. 
It typically requires the computation and rendering of a large number of particle trajectories, such as pathlines and finite-time Lyapunov exponents (FTLEs), which can be computationally expensive and memory intensive. 
Furthermore, these computational challenges limit the interactivity of flow visualization. 
To address these challenges, conventional  approaches decouple the particle advection and the rendering process to accelerate the  visualization performance. 
For instance, pathlines are pre-computed and then visualized by a texture-based approach~\cite{helgeland2006high} or distributed to a high-performance computing system~\cite{ali2010visualization}. 

Deep learning techniques are promising in addressing these computational challenges in time-varying flow visualization. 
They provide compact representations, have reduced memory footprints, and provide fast inference capabilities. 
Recent advancements have been in applying deep learning to various aspects of fluid dynamics~\cite{brunton2020machine}. 
Concurrently, the scientific visualization community has increasingly utilized deep learning in the visualization pipeline~\cite{liu2022deep, wang2022dl4scivis}, and specifically in the analysis and visualization of time-varying flow fields~\cite{han2021exploratory, sahoo2022neural, an2021stsrnet, han2022tsr}. 
Recently, Han et al.~\cite{han2021exploratory} provided a first step toward utilizing a deep learning approach for time-varying particle tracing. 
They employed a multi-layer perceptron (MLP) model to reconstruct Lagrangian-based flow maps. 
Whereas their results highlighted the advantages of scientific deep learning, such as reduced memory footprints  and efficient inference, their method was limited to a 2D analytic flow and lacked empirical evidence to showcase  the broader applicability of deep learning in time-varying flow visualization.  

\begin{figure}[]
    \centering
    \includegraphics[width=0.9\linewidth]{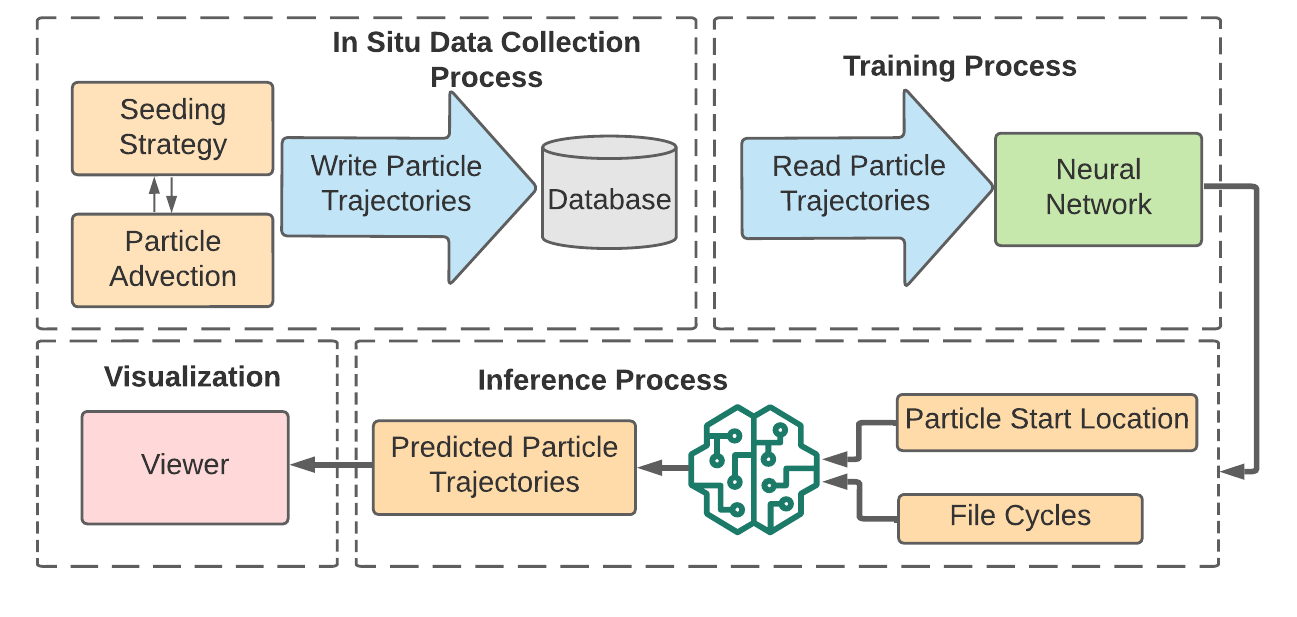}
    \vspace{-1em}
    \caption{The workflow of our deep learning-based particle tracing model. The Lagrangian flow maps are created using in situ processing, saved to the database, and input into a neural network to learn the corresponding end locations based on particle start locations and file cycles. Once the model has been fully trained, new particle trajectories can be inferred from the model and visualized using the developed viewer.\vspace{-2em}}
    \label{fig:workflow}
    
\end{figure}

In this paper, we provide an in-depth study of MLP-based particle tracing deep learning models in capturing Lagrangian representations of time-varying flows and demonstrate their capability to enable fast and accurate visualization of various flow regimes. 
The workflow of our deep learning approach is illustrated in \cref{fig:workflow}. 
We advance beyond the particle tracing model of Han et al.~\cite{han2021exploratory} by conducting a comprehensive evaluation of the MLP-based models compare with the conventional approach. 
Our long-term goal is to build a robust and efficient framework for Lagrangian-based flow visualization using deep learning.
To that end, we empirically establish best practices in designing MLP-based models for flow reconstruction. 
Our contributions include: 
\begin{itemize}[noitemsep]
\item We evaluate how effective MLP-based models are in reconstructing particle trajectories across a diverse collection of 2D and 3D flows, including structured and unstructured input data.
\item We perform an in-depth analysis of particle tracing MLP models that includes examining the effects of various activation functions, discerning the influence of model architectures, gauging the impact of flow complexity, and evaluating the effects of different training data structures.
\item We compare our models against prior particle tracing  methods that utilize flow maps for exploratory flow visualization, including the MLP model of Han et al.~\cite{han2021exploratory} and conventional interpolation techniques. 
\item We assess the practical performance of deploying our trained models through both web-based Javascript and high-performance C++ libraries. This evaluation offers an in-depth understanding of how the neural network performs in practice.
\item We investigate model pruning techniques that enhance inference efficiency without compromising accuracy by judiciously discarding nonessential model weights.
\item We introduce web-based and OSPRay-integrated viewers to assess the practical performance of our particle tracing models. Using Lagrangian-based flow representations, these tools provide interactive and seamless post hoc analysis and visualization of time-varying flow fields. 
\end{itemize}

\section{Related Work}
\label{sec:related-work}

\subsection{Lagrangian Flow Reconstruction and Visualization}
\label{sec:background_lagrangian_flow}

Eulerian and Lagrangian reference frames are used to represent time-dependent flow fields. 
Eulerian-based representations store the velocity fields directly and calculate particle trajectories by integrating the velocity fields.  
Lagrangian representations encode the flow behaviors using flow maps $F^t_{t_0}$, which store the particle start location and end location from time $t_0$ to time $t$ and calculate arbitrary particle trajectories using interpolation. 
Even though an Eulerian representation is fast to calculate, it requires a dense temporal resolution to obtain accurate trajectory reconstruction~\cite{da2004lagrangian,Qin2014,agranovsky2014improved,sane2018revisiting,rockwood2019practical,sane2021investigating}. 
In contrast, the Lagrangian-based representation has received increased attention as it provides good  accuracy-storage tradeoffs for exploration in temporally sparse settings~\cite{agranovsky2014improved, rapp2019void,sane2021investigating,sane2022exploratory}. 
It also directly supports feature extraction~\cite{froyland2015rough,schlueter2017coherent,hadjighasem2017critical,froyland2018robust,Jakob2020}. 

Using a Lagrangian representation, information is encoded with flow maps, computed using in situ processing, and analyzed post hoc. 
The reconstruction of new trajectories from the flow maps is a crucial component of post hoc analysis. 
Agranovsky et al. presented a multiresolution interpolation scheme that begins with a base resolution and adds additional trajectories if the region contains interesting behaviors~\cite{agranovsky2015multi}.  
Bujack et al.~\cite{bujack2015lagrangian} proposed representing particle trajectories by parametric  curves, such as B\'ezier curves and Hermite splines, to improve the aesthetics of the derived trajectories. 
Chandler et al.~\cite{chandler2014interpolation} developed a k-d tree for efficient lookup of the particle neighborhoods during interpolation. 
However, to the best of our knowledge, none of the existing works have investigated real-time exploration and visualization of Lagrangian-based flows. 
Two main challenges of interactive visualization during the post hoc analysis include reducing the I/O overhead of loading high-resolution flow maps and accelerating the cell lookup of the particle neighborhoods. 
In addition, unstructured flow maps require time-consuming triangulation or tetrahedralization, making the interpolation process even slower. 

In this paper, we empirically evaluate the use of deep learning for post hoc reconstruction and demonstrate the framework using an interactive web-based viewer for visualizing and analyzing the flow field. 
Using deep learning, flow maps can be represented by a model to conserve storage space. 
Importantly, interactive queries for arbitrary particle trajectories are possible without requiring intensive I/O operations for loading flow maps or performing the cell lookup procedure.

\subsection{Deep Learning for Flow Visualization}
\label{sec:background_dl4flowvis}

Deep learning methods have become increasingly popular for flow visualization~\cite{liu2022deep}. 
Examples of their widespread applications include the detection of eddies and vortices~\cite{lguensat2018eddynet, yi2018cnn, strofer2018data, bai2019streampath, duo2019oceanic, liu2019cnn,deng2019cnn, wang2021rapid,franz2018ocean, kashir2021application,beck2020neural,tatarenkova2020edge}, the segmentation of streamlines~\cite{li2015extracting}, the extraction of a stable reference frame from unsteady 2D vector fields~\cite{kim2019robust}, the optimization of data access patterns to boost computational performance in distributed memory particle advection~\cite{hong2018access}, and the selection of a representative set of particle trajectories~\cite{sane2020survey} using clustering methods grounded in deep learning~\cite{han2018flownet, lee2021deep}. 
Furthermore, data reduction and reconstruction is  widely discussed, due to the scale of the flow data. 
Recent works have used low-resolution data~\cite{guo2020ssr,gao2021super,hohlein2020comparative} or 3D streamlines~\cite{han2019flow, sahoo2021integration} to reconstruct high-resolution flow fields. 
Using an efficient subpixel convolutional neural network (ESPCN)~\cite{shi2016real} and a super-resolution convolutional neural network (SRCNN)~\cite{dong2015image}, Jakob et al.~\cite{Jakob2020} upsampled 2D FTLE scalar fields produced from Lagrangian flow maps.
Sahoo et al.~\cite{sahoo2022neural} proposed a reconstruction technique for compressing time-varying flow fields using implicit neural networks. 

Successfully visualizing flow map data relies on two critical factors: (1) accurate reconstruction of particle trajectories and (2) interactive visualization and exploration of these trajectories. Han et al.~\cite{han2021exploratory} employed a MLP architecture to reconstruct Lagrangian-based flow maps for a 2D analytic dataset. 
Whereas they demonstrated the accuracy of reconstructing flow fields using a neural network, their study lacked quantitative and qualitative assessments across various datasets and an  exploration of potential model architecture modifications. 
Furthermore, they did not investigate the benefits of rapid inference provided by deep learning models for interactive visualization.

Our research utilizes the exact Lagrangian representation of time-varying flow fields as data for neural networks, constructed using MLP and sinusoidal activation functions~\cite{sitzmann2020implicit}. We assess our method through qualitative and quantitative analyses on a variety of 2D and 3D datasets, including both structured and unstructured input. We conduct comprehensive experiments to investigate the impact of model architecture, flow complexity, and activation function. We also compare the performance of our deep-learning-based approach with the conventional post hoc interpolation method.
Moreover, we utilize deep learning for interactive visualization and exploration during post hoc analysis of Lagrangian-based flow fields. 
We demonstrate that the neural network can be seamlessly integrated with various rendering APIs written in different programming languages and deployed on different platforms.


\section{Lagrangian Analysis Using Deep Learning}
\label{sec:method-network}

Our neural network is designed to learn the behavior of a  time-varying flow field. 
Before model training, we compute the Lagrangian flow maps by advecting massless particles in a time-varying flow field to generate the training datasets (\cref{sec:method_data_gen}).
We adapt the MLP network architecture from Han et al.~\cite{han2021exploratory}, which consists of an encoder and a decoder built with a series of fully connected (FC) layers. 
The starting location of a seed and a file cycle are individually input into encoders equipped with FC layers. 
The resulting encoded latent vectors are then concatenated and fed into the decoding layers, which predict the seed's end location at the specified file cycle. 
For comparative analysis, our neural network uses the same number of FC layers for the encoders of the seeds' start location and file cycles to  investigate the impact of the number of layers and the size of the latent vector on the performance. 
Since Sitzmann et al.~\cite{sitzmann2020implicit} showed that sinusoidal activation function is suited for representing complex natural signals, we replace the ReLU activation function with the sinusoidal activation function (\cref{sec:method_model}).

\subsection{Training Data Generation}
\label{sec:method_data_gen}

\begin{figure}[]
    \centering
    \includegraphics[width=\linewidth]{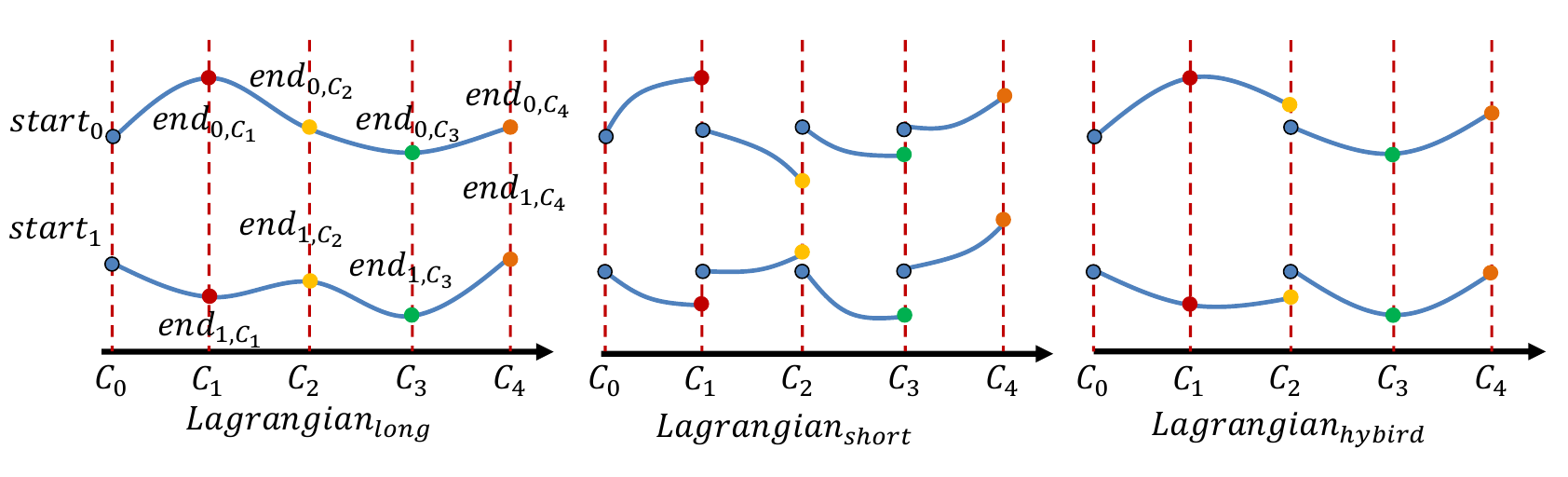}
    \vspace{-2em}
    \caption{Illustration for $Lagrangian_{long}$, $Lagrangian_{short}$ and $Lagrangian_{hybird}$ methods  using 1D particle trajectories. The x-axis represents the file cycle. Circles with the same color have the same labels. The $Lagrangian_{long}$ extracts a single flow map with end locations at uniform time intervals along the trajectories. In contrast, the $Lagrangian_{short}$ extracts several short flow maps, resetting start seeds for each time interval. The $Lagrangian_{hybird}$ combines the strengths of both $Lagrangian_{long}$ and $Lagrangian_{short}$: it extracts the $Lagrangian_{short}$ flow maps, where each individual flow map follows the structure of a $Lagrangian_{long}$ flow map. $Lagrangian_{hybird}$ achieves both comprehensive domain coverage and comparable accuracy.\vspace{-2em}}
    \label{fig:lagrangian_appraoch}
\end{figure}

In this study, we replicate the training data generation of Han et al.~\cite{han2021exploratory}, which uses two methods to extract flow maps --- $Lagragian_{long}$ and $Lagrangian_{short}$ (\cref{fig:lagrangian_appraoch}). 
The $Lagrangian_{long}$ method extracts a single flow map composed of long particle trajectories with uniform temporal sampling for each integral curve.
In contrast, the $Lagrangian_{short}$ method extracts multiple short flow maps, each comprising a set of seed locations and a set of end locations for each seed, where each end location corresponds to the displacement from the start location over nonoverlapping intervals. 
Each method has its advantages and disadvantages.
$Lagrangian_{short}$ flow maps provide good domain coverage since particles are periodically reset, but they may incur error propagation and accumulation when deriving new particle trajectories. 
$Lagrangian_{long}$ flow maps enable the derivation of new trajectories free of error propagation. 
However, as the integration time increases, the domain coverage of this method deteriorates and interpolation accuracy may decrease as particles diverge. 
To leverage the benefits of both methods, we introduce a hybrid method called $Lagrangian_{hybrid}$ (\cref{fig:lagrangian_appraoch}). 
This method extracts multiple $Lagrangian_{short}$ flow maps, where each map itself is a $Lagrangian_{long}$ flow map composed of particle trajectories with uniform temporal sampling for each integral curve instead of just storing the end locations. 
The training data structure is consistent across all three methods. 
However, there are slight differences in the inference processes for each method. 
We explore these differences and provide an accuracy comparison in \cref{sec:num_fm}.

For the seeding method, we place the initial seeds using a Sobol quasirandom sequence (Sobol), which has performed better than the pseudorandom number sequence and uniform grid~\cite{han2021exploratory}. 
The initial step in the production of training data is the placement of seeds in the spatial domain. 
After the placement of the seeds, particle trajectories are determined by shifting particles from $t$ to $t + \delta$, where $\delta$ represents one simulation time step. 
We refer to one simulation time step as a \emph{cycle}. 
The cycle on which the end locations are saved is a \emph{file cycle}, and the number of cycles between two successive file cycles is an \emph{interval}. 

Given a total temporal duration of $T$, seeds are inserted once at the beginning of time $t_0$ and traced until $T$ to produce flow maps using the $Lagrangian_{long}$ method. 
During the particle tracing process, intermediate locations are saved. Using the $Lagrangian_{short}$ method, particle tracing begins at time $t_0$ and concludes at time $t_1 = t_0 + \delta \times I$, where $I$ is the interval.
The location at $t_1$ is then recorded, and the tracing seeds are reset until the next file cycle. This process is repeated until the last file cycle.

The $Lagrangian_{hybrid}$ method also begins particle tracing at time $t_0$ and terminates at $t_1 = t_0 + \delta \times I  \times p $, where $I$ is the interval between the file cycle and $p$ is the number of intermediate locations to trace.
The intermediate locations between $(t_0, t_1]$ are recorded, and at time $t_1$, the seeds are reset. This process is continued until the final file cycle. The datasets used for training are saved in the NPY file format for efficient Python loading. 

We built an $m \times n \times n$ array to store seed start locations and end locations across file cycles, where $m$ represents the number of seeds and $n$ represents the number of flow maps (file cycles). 
These training samples are arranged according to \cref{eqn:input}. Each training sample includes a start location $s_i$, the file cycle $c_j$ incorporating temporal data, and the target end location at the corresponding file cycle $\ell_{i,j}$ (where $0 \leq i \leq m-1$, $0 \leq j \leq n-1$).
The training dataset for our model is therefore 
\begin{equation}
    \label{eqn:input}
    \begin{aligned}
    Input = & \{\{s_0, \,c_0, \,\ell_{0, 0}\}, 
              \{s_0, \,c_1, \,\ell_{0, 1}\}, ..., \\
             & \{s_0, \,c_{n-1}, \,\ell_{0, {n-1}}\}, ..., 
              \{s_{m-1}, \,c_{n-1}, \,\ell_{m-1, {n-1}}\}\}. 
    \end{aligned}
\end{equation}

\subsection{Network Architecture}
\label{sec:method_model}
Our neural network is based on the MLP architecture developed by Han et al.~\cite{han2021exploratory}. 
The encoder \textbf{E} takes the particle start locations coupled with file cycles as input and passes them through two sequences of FC layers that are then concatenated to form the latent vector input for the decoder \textbf{D}. 
The decoder \textbf{D} outputs the predicted end location, which is compared to the desired end location to calculate the loss. 
Although there is no universal model architecture suitable for all datasets, MLP offers flexibility in adjusting the size of the hidden vector and the number of layers to suit different datasets. 
In \cref{sec:model_selection}, we examine how changing the number of layers and the dimension of the hidden vector affects the model performance. 
Additionally, Han et al.~\cite{han2021exploratory} observed that reconstruction errors increase as the number of file cycles (learned by a model) increases. 
As flow maps store nonoverlapping intervals, we train multiple models by partitioning the flow maps. 
However, using multiple models involves a trade-off between memory consumption and precision, which we investigate in \cref{sec:num_fm}.

\begin{figure}[]
    \centering
    \includegraphics[width=\linewidth]{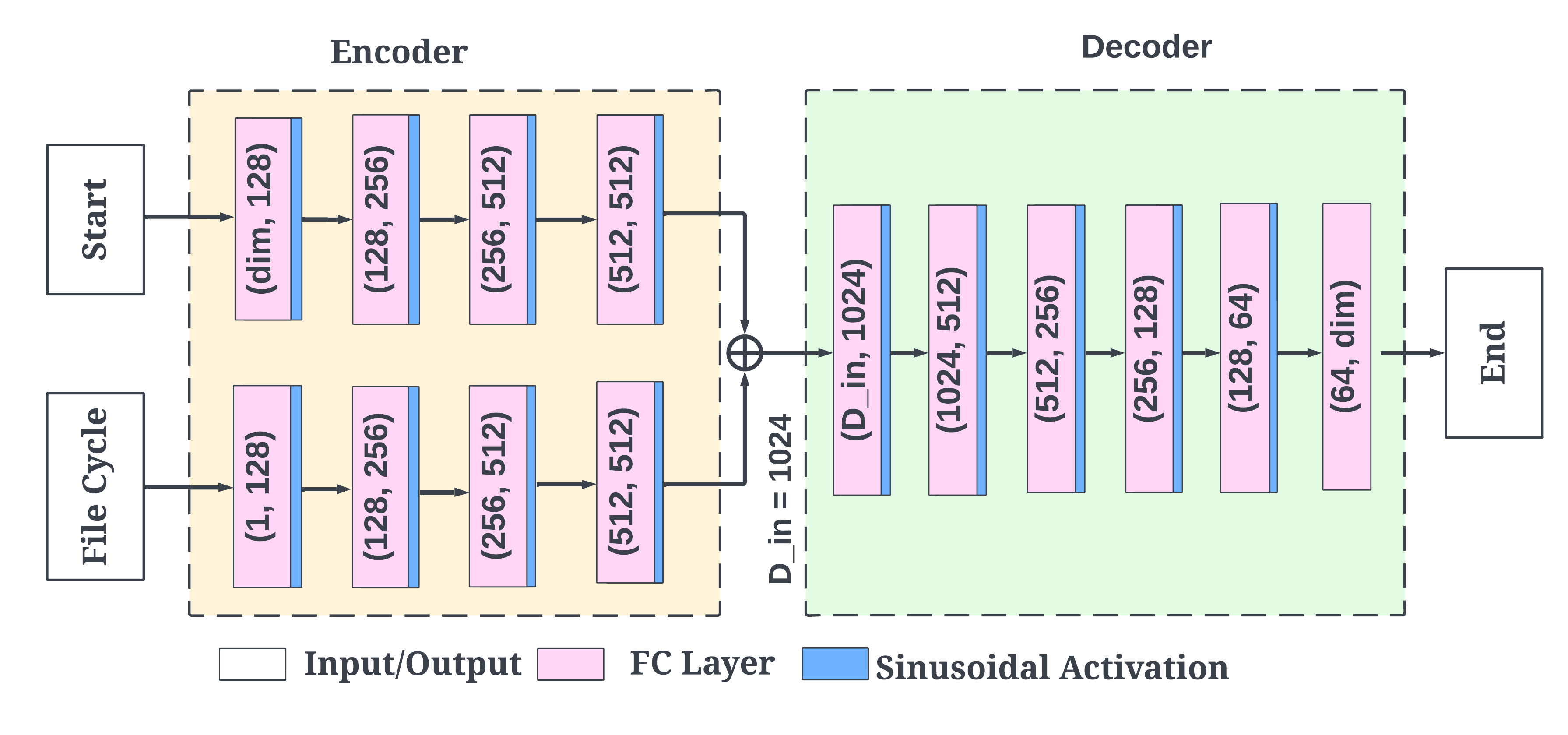}
    \vspace{-2.3em}
    \caption{The MLP architecture of our neural network. The network architecture begins by taking two inputs: the particle's initial location (Start) and the number of file cycles (File Cycle). These inputs are first processed by the Encoder, which transforms them into a latent vector represented as $D_{in}$. Following this, the latent vector $D_{in}$ is input into the Decoder. The Decoder then processes this information to output the final location (End) of the particle at the queried file cycle. The sinusoidal activation function is used after the FC layers in the model except the output layer.\vspace{-2em}}
    \label{fig:models}
\end{figure}

The neural network is developed using Pytorch\footnote{https://pytorch.org}. 
We use the Adam optimizer with the hyperparameters $\beta_1 = 0.9$, $\beta_2 = 0.999$, and $\epsilon = 1e^{-6}$ along with a learning rate scheduler to reduce the current learning rate by a factor of 2 if the validation loss has not dropped for 5 epochs. 
We utilize an L1 loss as our loss function:
\vspace{-0.5em}
\begin{equation}
    \label{eqn:loss}
    \vspace{-0.5em}
    \begin{aligned}
    loss(\ell_{i, j}, \hat{\ell}_{i, j}) = L1Loss(\ell_{i, j}, \hat{\ell}_{i, j}), 
    \end{aligned}
\end{equation}
where $\ell_{i, j}$ represents the target (ground truth) end location of seed $i$ at file cycle $c_j$ and $\hat{\ell}_{i, j}$ denotes the predicted end location ($0 \leq i \leq m-1$ and $0 \leq j \leq n-1$).

\section{Results and Discussion} 
\label{sec:results}

We first describe the datasets used in our evaluation (\cref{sec:data_sets}).
To investigate the best practices involving deep-learning-based particle tracing, we then evaluate the impact of model architecture, activation function, flow field complexity, and the training data structure on the performance of our model (\cref{sec:model_evaluation}). 
Additionally, we discuss model pruning to reduce the size of trainable parameters and evaluate the inference performance of our proposed neural network deployed by web-based JavaScript and high-performance C++ application (\cref{sec:model_pruning}). 
We also conduct a comparative analysis against prior techniques that employ flow maps as input for exploratory flow visualization, such as the predecessor model of Han et al.\cite{han2021exploratory} (\cref{sec:activation})  and the conventional Barycentric coordinate interpolation method (\cref{sec:comparison}). 
Finally, we introduce a web-based viewer and an OSPRay-based viewer, offering two deployment options for the trained model and facilitating interactive flow visualization and exploration (\cref{sec:viewer}). 
Our experiments employ a Dual RTX 3090s GPU for model training in a dual-socket workstation with two Intel Xeon E5-2640 v4 CPUs (40 logical cores at 2.4 GHz and 128 GB RAM) and an NVIDIA Titan RTX GPU for evaluation.

\subsection{Datasets}
\label{sec:data_sets}

In our studies, we utilize seven datasets, including four 3D datasets: ABC, Structured/Unstructured Half Cylinder ensembles, ScalarFlow, and Hurricane; and three 2D datasets: Double Gyre, Gerris Flow ensembles, and Structured/Unstructured Heated Cylinder.

Standard benchmark datasets such as the \textbf{Double Gyre} and \textbf{ABC} are commonly employed in fluid dynamics research, specifically for developing flow visualization techniques and tools. 
The Double Gyre flow field is defined within the spatial domain of $[0, 2] \times [0, 1]$, while the ABC flow field is defined within the spatial domain of $[0, 2\pi] \times [0, 2\pi] \times [0, 2\pi]$. The equations used for the simulations are available in the supplemental material.

\textbf{Heated Cylinder} is a 2D unsteady simulation generated by a heated cylinder with Boussinesq Approximation~\cite{gerrisflowsolver, Guenther17}. 
The simulation domain is $[-0.5, 0.5] \times [-0.5, 2.5] \times [0, 20]$. 
Our experiments utilize the time span from 0 to 10. 
To demonstrate our approach, we employ both structured and unstructured datasets. 
The structured dataset has a grid resolution of $150 \times 450$.

\textbf{Gerris Flow} is a 2D ensemble simulation generated by a Gerris flow solver~\cite{gerrisflowsolver, Jakob2020}.  
It contains 8000 datasets with the value of Reynolds number (Re) varying from a steady regime ($Re < 50$) to periodic vortex shedding ($Re < 200$) to turbulent flows ($Re > 2000$)~\cite{Jakob2020}. We choose datasets with a $Re$ value of $23.2$, $101.6$, $445.7$, and $2352.5$, respectively, to showcase the performance of our method for varying degrees of flow complexity.
The grid resolution $[X \times Y \times T] = [512 \times 512 \times 1001]$ with a simulation domain of $[0, 1] \times [0, 1] \times [0, 10]$.

\textbf{Half Cylinder} is a 3D ensemble of numerical simulations of an incompressible 3D flow around a half cylinder~\cite{gerrisflowsolver, BaezaRojo19SciVisa}. 
We experiment with both structured and unstructured grids, selecting $Re$ values of $160$ and $320$ to investigate the effects of varying turbulence degrees. 
The domain of the simulation is set to be $[-0.5, 7.5] \times [-0.5, 1.5] \times [-0.5, 0.5]$.
Both structured and unstructured datasets span $80$ time steps.
The structured dataset has a grid resolution of $640 \times 240 \times 80$.

\textbf{ScalarFlow} is a large-scale, 3D reconstruction of real-world smoke plumes~\cite{eckert2019scalarflow}. 
The spatial dimension is $[100 \times 178 \times 100]$, and the number of time steps is $150$.

\textbf{Hurricane} is a simulation from the National Center for Atmospheric Research\footnote{http://vis.computer.org/vis2004contest/data.html}.
The data dimension is $[500 \times 500 \times 100]$ with 47 time steps. 
We use the first $40$ time steps and the region $[150, 399] \times [150, 399] \times [0, 99]$ that contains the interesting feature---the hurricane eye.

In our experiments, we use a step size $\delta = 0.01$ and interval $I=5$ for the Double Gyre, ABC, Unstructured Heated Cylinder, and Gerris Flow datasets. We set  $\delta = 0.1$ and $I = 1$ for the Half Cylinder, ScalarFlow and Hurricane datasets.

\subsection{Model Evaluation\vspace{-0.5em}}
\label{sec:model_evaluation}

We first investigate the effect of model architecture  parameters on performance, such as the number of layers and the size of the hidden vector (\cref{sec:model_selection}). 
Next, we compare our model using the sinusoidal activation function with previous work~\cite{han2021exploratory} that uses the ReLU activation function to demonstrate the advantages of the sinusoidal activation  (\cref{sec:activation}). 
Finally, we illustrate the performance of the neural network on the flow complexity using ensemble datasets and enhance the accuracy by training multiple models along the pathlines and examining the effectiveness of our $Lagrangian_{hybird}$ method (\cref{sec:num_fm}). 

During the training phase, we generate an additional 10\% of training data samples for validation. 
In the following testing results, each error instance represents the average distance between the predicted and target (ground truth) locations along a trajectory, as defined in \cref{eqn:error}.  
\vspace{-1em}
\begin{equation}
\vspace{-0.8em}
\label{eqn:error}
    error_i = \frac{1}{n}\sum_{j=0}^{n-1}loss(\ell_{i, j},  \hat{\ell}_{i, j})
\end{equation}
where $i$ represents the index of the new seed and $n$ is the number of end locations (file cycles)
along the trajectories. $\ell_{i,j}$ is the target (ground truth) end location, and the $\hat{\ell}_{i,j}$ is the predicted end location. 
We place $5,000$ random seeds for all testing results presented below. 

\subsubsection{Impact of Model Architecture}
\label{sec:model_selection}

We evaluate the qualitative and quantitative effects of varying the number of layers in the encoder and decoder as well as the dimension of the encoded latent vector. 
We use the Double Gyre, Gerris Flow with $Re = 101.6$ and $Re = 445.7$, ABC flow, and the unstructured Half Cylinder with $Re = 160$ and $Re = 320$. 
For benchmarking model performance, we utilize $100$ flow maps for each 2D dataset and $50$ flow maps for each 3D dataset. 
For all datasets, the number of seeds we distribute is half of the grid resolution; except in the case of the Half Cylinder dataset, we strategically place seeds only within the region defined by $[-0.5, 0.5] \times [-0.5, 0.5] \times [-0.5, 0.5]$ to optimize training time. 
This region encompasses the obstruction and areas of interest, allowing for more focused training. 

In the experiments, we set the number of decoder layers and encoder layers to be four, six, and eight, respectively. 
The encoded latent vector has dimensions of 1024 and 2048.  
Tab.~1 in the supplemental material displays the maximum, mean, and median errors associated with various combinations of encoder layers, decoder layers, and latent vector dimensions.

\begin{table}[H]
\centering
\resizebox{0.8\columnwidth}{!}{
\begin{tabular}{ccccccc}
\hline
& \multicolumn{2}{c}{Model(MB)} & \multicolumn{2}{c}{Training (hrs)} & \multicolumn{2}{c}{Inference (s)} \\ 
\cmidrule(rl){2-3} \cmidrule(rl){4-5} \cmidrule(rl){6-7}
{[}\#E, \#D{]} & 1024          & 2048          & 1024             & 2048            & 1024            & 2048            \\ \cmidrule(rl){1-1} \cmidrule(rl){2-3} \cmidrule(rl){4-5} \cmidrule(rl){6-7}
{[}4, 4{]}     & 10.246           & 40.939          & 0.348            & 0.478           & 0.297           & 0.343           \\
{[}4, 6{]}     & 10.409           & 41.592          & 0.368            & 0.488           & 0.277           & 0.397           \\
{[}4, 8{]}     & 10.419           & 41.633          & 0.380            & 0.499           & 0.308           & 0.351           \\
{[}6, 4{]}     & 10.327           & 41.265          & 0.379            & 0.491           & 0.282           & 0.375           \\
{[}6, 6{]}     & 10.490           & 41.919          & 0.396            & 0.502           & 0.285           & 0.360           \\
{[}6, 8{]}     & 10.500           & 41.960          & 0.410            & 0.511           & 0.279           & 0.354           \\
{[}8, 4{]}     & 10.332           & 41.286          & 0.409            & 0.502           & 0.329           & 0.359           \\
{[}8, 6{]}     & 10.495           & 41,940          & 0.425            & 0.512           & 0.283           & 0.376           \\
{[}8, 8{]}     & 10.505           & 41.980          & 0.438            & 0.521           & 0.289           & 0.366           \\ \hline
\end{tabular}
}
\caption{\label{table:model_size} The training and inference time for models of varying sizes for the Gerris Flow ($Re=445.7$) dataset. The model size, along with the training and the inference time, is not affected by the dataset. Each row ($[E, D]$) represents the number of encoding layers ($E$) and decoding layers ($D$), and the hidden vector dimension is either $1024$ or $2048$. Training time is measured on $100K$ training samples with $20$ flow maps, trained for $100$ epochs, utilizing distributed data parallel. Inference time was evaluated on $5,000$ seeds with $20$ flow maps, using Pytorch in Python with CUDA.\vspace{-1em}}
\end{table}

Our model design achieves an acceptable error rate for all datasets. 
Our experiments reveal that the optimal model architecture depends on the specific dataset. For steady flow regimes, such as those observed in the Double Gyre and Gerris ($Re=23.2$), a smaller latent vector dimension results in a higher accuracy. 
Conversely, for datasets exhibiting periodic vortex shedding, such as Gerris ($Re=101.6$) and some 3D datasets, a larger latent vector dimension is shown to be more effective than a smaller one. 
Furthermore, we observe a trend regarding the depth of encoding and decoding layers. In most cases, models with deeper layers, particularly those with eight encoding or decoding layers, yield less accurate inferences, performing the least effectively in our tests (see Tab.~1 in the supplemental material). 
When evaluating a neural network, model size is an important consideration, in addition to its inference accuracy. 
As shown in \cref{table:model_size}, model size increases with an increasing number of layers in both the encoder and decoder, and doubles when increasing the hidden vector size from $1024$ to $2048$. 
Moreover, the neural network with more parameters requires more time for both training and inference.

\subsubsection{Impact of Activation Function}
\label{sec:activation}

We compare our approach with the ReLU-based MLP model proposed by Han et al.\cite{han2021exploratory} with the same model architecture, which includes five encoding layers for the seeds' start location, seven encoding layers for the file cycles, a latent vector of dimension $1024$, and six decoding layers. 
As recommended by Sitzmann et al.~\cite{sitzmann2020implicit}, we remove the LayerNorm and replace the ReLU activation function in the original architecture with the sinusoidal function for our comparative  study. 
For our experiments, we use the same datasets as described in \cref{sec:model_selection}. 
We set the learning rate to be $1e^{-4}$ for the ReLU-based MLP, as demonstrated in \cite{han2021exploratory} to be the optimal choice and confirmed by our experiments. 
For our approach using the sinusoidal function, we use an optimal learning rate of $5e^{-4}$.

\begin{figure}[!ht]
    \centering
    \includegraphics[width=0.9\linewidth, keepaspectratio]{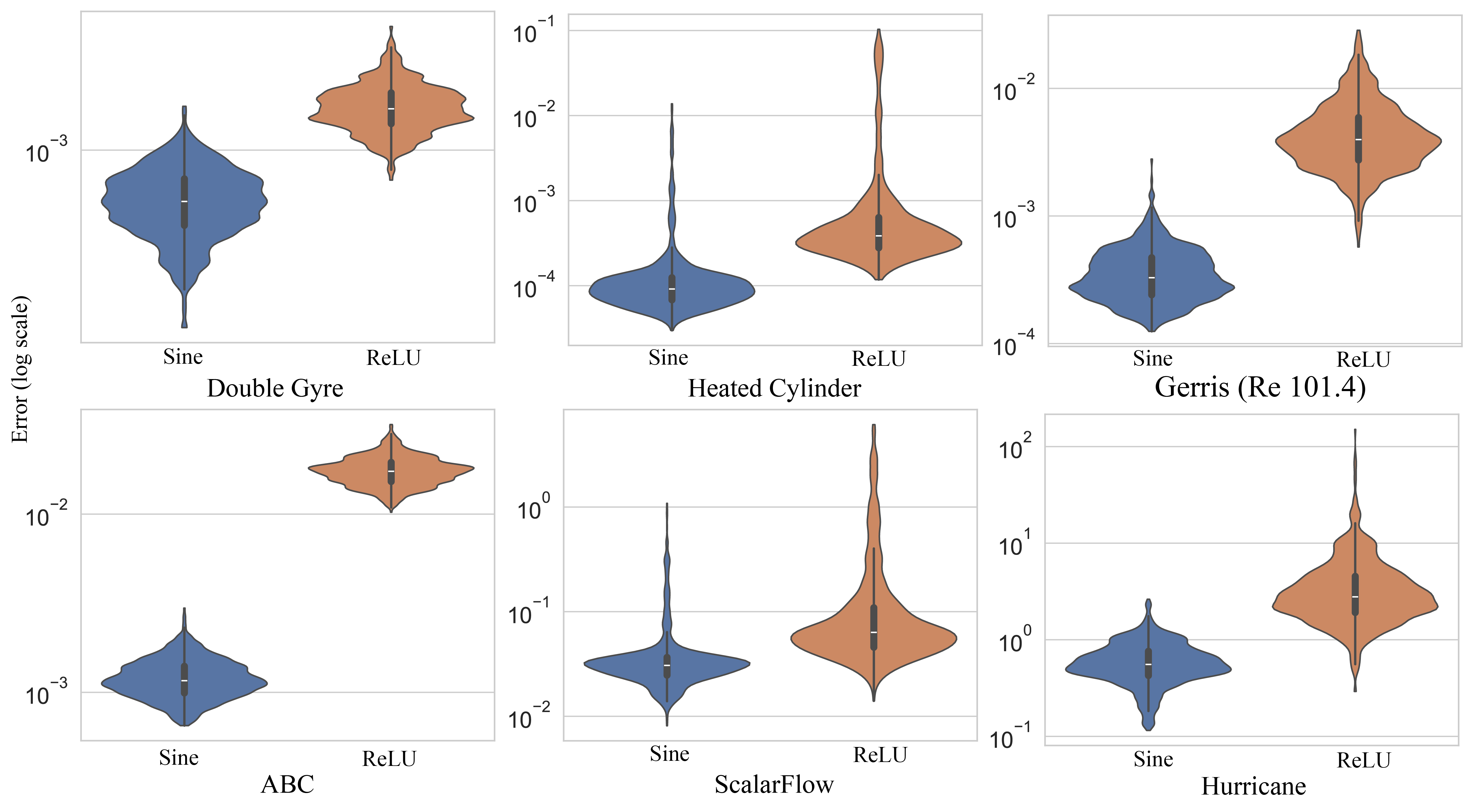}
    \caption{The error plot compares our sinusodial-based MLP  with the ReLU-based MLP of Han et al.~\cite{han2021exploratory}. The sinusoidal activation function improves the inference accuracy significantly.\vspace{-2em}}
    \label{fig:activation_function_errors}
\end{figure}

\cref{fig:activation_function_errors} shows the inference error of our sinusoidal-based MLP and the ReLU-based MLP. 
Our results indicate that the sinusoidal activation function significantly outperform the ReLU activation function on all datasets tested. 
Furthermore, our method achieves these results using the same storage space as the model in~\cite{han2021exploratory} (which has a size of $8.4$  MB), while greatly improving the accuracy.

\subsubsection{Impact of Flow Map Extraction Strategies}
\label{sec:num_fm}

The optimal model architecture varies among datasets. 
Based on these findings, we investigate how the complexity of the flow affects the performance of our deep learning model. 
We conduct an analysis on four Gerris Flow datasets with varying $Re$ values: $23.2$, $101.6$, $445.7$, and $2352.5$, respectively, covering a range of flow regimes from steady ($Re < 50$), periodic vortex shedding ($50 < Re < 200$), to turbulent flows ($Re > 2000$)~\cite{Jakob2020}. 
Each training dataset consists of $500$ time steps with an interval of $5$, resulting in $100$ flow maps for each dataset. 
To train our model, we use $262,144$ seeds for each dataset and design the model with four encoding layers, six decoding layers, and a $2048$D latent vector.

\begin{figure}[!htb]
    \centering
    \vspace{-1em}
    \includegraphics[width=0.9\linewidth, keepaspectratio]
    {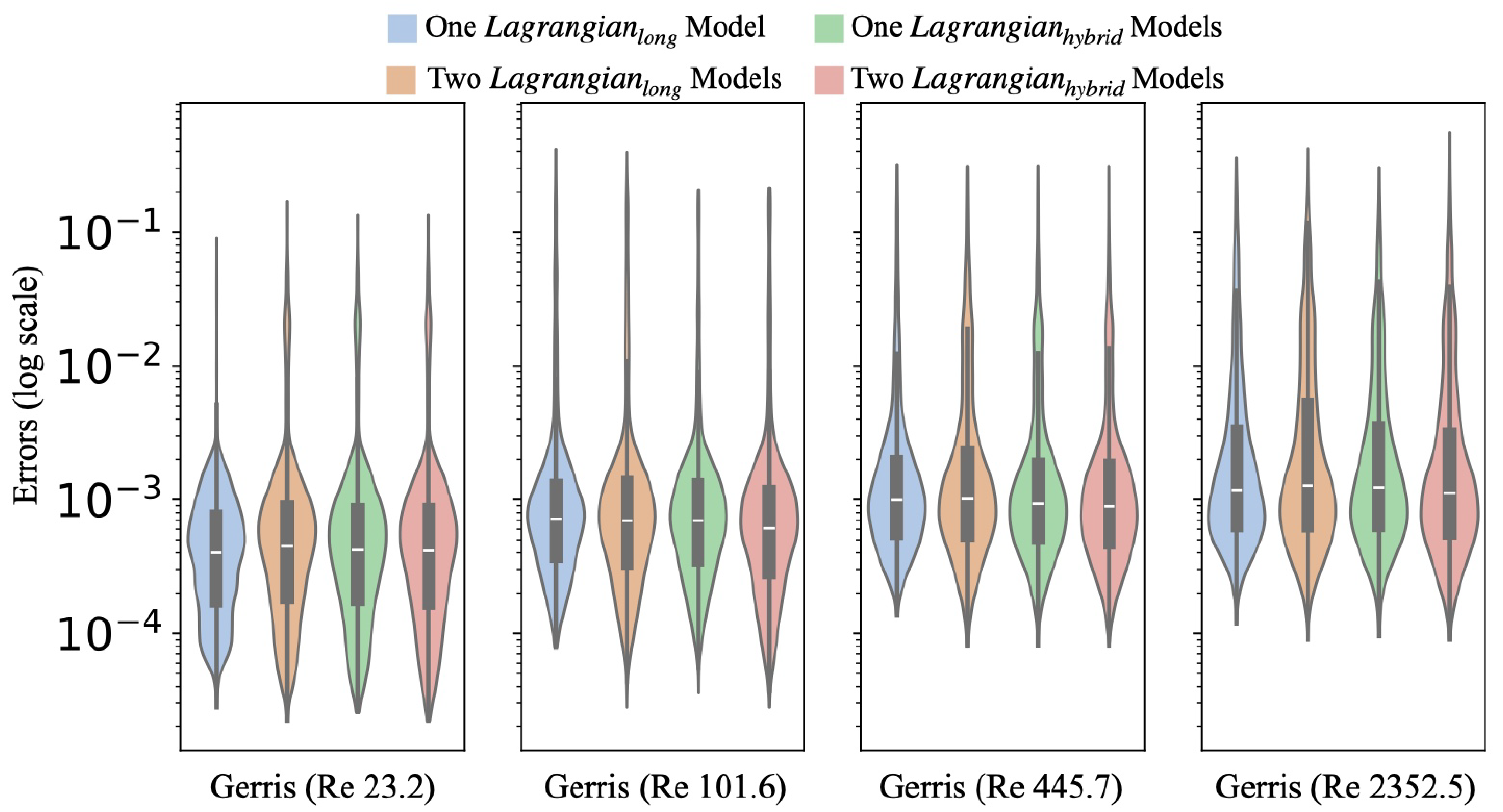}
    \vspace{-.7em}
    \caption{Violin plot depicting the inference errors for models trained on Gerris Flow dataset with varying Reynolds numbers, using either the $Lagrangian_{long}$ method with one or two models or the $Lagrangian_{hybrid}$ method with one or two models. These evaluations, conducted with $5,000$ testing seeds, calculate errors following \cref{eqn:error}. A comparison of violin plots of the same color across different datasets reveals that errors tend to increase as the flow behavior becomes more turbulent. Interestingly, using fewer flow maps for training in the $Lagrangian_{long}$ approach does not improve accuracy. Conversely, in the $Lagrangian_{hybrid}$ approach, using fewer flow maps actually led to a slight decrease in median errors. Furthermore, the $Lagrangian_{hybrid}$ method effectively reduces error propagation while maintaining domain coverage.\vspace{-2em}}
    \label{fig:re_errors}
\end{figure}

Our results reveal that one of the limitations of our  method is the decline in inference accuracy as the underlying flow behavior becomes more complex. Specifically, we observe that the turbulence flow feature (Gerris with $Re = 2352.5$) is the most challenging to learn (\cref{fig:re_errors}). 
Improving the ability to infer turbulent flow could be a focus of future research. 

Prior work has demonstrated the $Lagrangian_{long}$ method results in fewer errors compared to $Lagrangian_{short}$, when using conventional~\cite{sane2019interpolation} or deep learning~\cite{han2021exploratory} methods, despite a reduction in domain coverage over time. 
Consequently, our experiments focus solely on comparing the $Lagrangian_{long}$ and $Lagrangian_{hybrid}$ methods  (\cref{fig:lagrangian_appraoch}).
As shown in \cref{fig:re_errors}, the $Lagrangian_{hybrid}$ method consistently reduces median errors across all datasets when compared to the $Lagrangian_{long}$ method. 
In addition, our study explores the effect of reduction on the number of flow maps on training accuracy. This is achieved by training two separate models, each with $50$ flow maps, as opposed to a single model trained with $100$ flow maps.

The findings, as depicted in \cref{fig:re_errors}, indicate that reducing the number of flow maps per model does not improve training accuracy when employing the $Lagrangian_{long}$ method. However, we observe a slight decrease in errors  when implementing the $Lagrangian_{hybrid}$ method. 
Throughout the training phase, the second model employing the $Lagrangian_{long}$ method exhibits errors approximately five times greater than those of the first model. 
We hypothesize that these increased errors may be due to the greater difficulty in achieving model convergence, possibly attributed to the larger spatial gap between the initial seed locations and their corresponding end locations in the second model.

Our findings suggest that $Lagrangian_{hybrid}$ has greater accuracy than $Lagrangian_{long}$, and can also mitigate the error propagation issue of $Lagrangian_{short}$.

\subsection{Model Pruning and Inference Performance}
\label{sec:model_pruning}

\subsubsection{Model Pruning}
An MLP model can often contain more weights than necessary for a given task. 
The lottery ticket hypothesis suggests that a small subset of the network is responsible for the majority of the task~\cite{frankle2018lottery}. 
This hypothesis implies that a small model within a large one can achieve the same level of accuracy. 
In order to create more efficient neural networks, we utilize model pruning techniques, specifically the algorithm proposed by Fang et al.~\cite{fang2023depgraph} using DepGraph. 
This approach uses a magnitude pruning technique~\cite{elesedy2020lottery} that automatically identifies the most important connections in the neural network, resulting in a smaller, more efficient model. 
We perform the pruning interactively by removing some weights and then fine-tuning the model for each epoch; we repeat the process until a certain percentage of the model is pruned. 
We do not prune important layers, such as the last layer of the model, to ensure that the model output retains the same format.
Utilizing structured pruning techniques allows us to create smaller, more efficient models that maintain the same level of accuracy as their larger counterparts. 

A prime example of model pruning is demonstrated in our experiments with the Hurricane dataset, where we are able to reduce the size of a model from $34.5$ MB to just $13.7$ MB while maintaining the same level of accuracy. 
This reduction in model size can lead to faster inference times, decreased memory requirements, and improved overall efficiency in neural network applications. 

\subsubsection{Inference Performance}
In \cref{sec:model_selection}, we evaluate the inference speed of our models with different sizes using Pytorch in Python. 
In this section, we evaluate the inference performance of our neural network in the web-based viewer and OSPRay integration using the ABC and Hurricane datasets. 
For the ABC dataset, we build a model with a $1024$D hidden vector, three encoding layers, and six decoding layers, resulting in a model size of $8.4$ MB.  
For the Hurricane dataset, we construct a model with a $2048$D hidden vector, five encoding layers, and eight decoding layers. The model size is $13.7$ MB after pruning.
The ABC dataset contains $20$ flow maps, and the Hurricane dataset contains $30$ flow maps.
%
%
We evaluate the performance of our viewers on two workstations: one with an Intel Xeon CPU and an NVIDIA Titan RTX GPU, and the other with an Intel NUC i7-8809G CPU (8 logical cores and 32 GB RAM). We measure the speed of rendering by scattering 100, 200, 300,
400, 500, and 1000 seeds across the domain for each dataset.


\begin{table}
\begin{subtable}{\columnwidth}\centering
{\resizebox{\columnwidth}{!}{
\begin{tabular}{cccccccc}
\hline
\multicolumn{8}{c}{ABC - $Lagrangian_{long}$ - 8.4 MB}                      \\ 
\multicolumn{1}{c}{} & \multicolumn{1}{c}{Model Loading (s)} & \multicolumn{1}{c}{\#100 (s)} & \multicolumn{1}{c}{\#200 (s)} & \multicolumn{1}{c}{\#300 (s)} & \multicolumn{1}{c}{\#400 (s)} & \multicolumn{1}{c}{\#500 (s)} & \#1000 (s)
\\ 
\cmidrule(rl){2-2} \cmidrule(rl){3-8} 

\multicolumn{1}{c}{ONNX + GPU} & \multicolumn{1}{c}{2.12}          & \multicolumn{1}{c}{0.45}  & \multicolumn{1}{c}{0.52}  & \multicolumn{1}{c}{0.58}  & \multicolumn{1}{c}{0.64}  & \multicolumn{1}{c}{0.65}  & 1.07   \\ 

\multicolumn{1}{c}{ONNX + CPU} & \multicolumn{1}{c}{1.69}          & \multicolumn{1}{c}{0.39}  & \multicolumn{1}{c}{0.82}  & \multicolumn{1}{c}{1.06}  & \multicolumn{1}{c}{1.41}  & \multicolumn{1}{c}{1.84}  & 3.19   
\\ \hline

\multicolumn{8}{c}{Hurricane - $Lagrangian_{long}$ - 13.7 MB}               \\ 
\multicolumn{1}{c}{} & \multicolumn{1}{c}{Model Loading (s)} & \multicolumn{1}{c}{\#100 (s)} & \multicolumn{1}{c}{\#200 (s)} & \multicolumn{1}{c}{\#300 (s)} & \multicolumn{1}{c}{\#400 (s)} & \multicolumn{1}{c}{\#500 (s)} & \#1000 (s)
\\ 
\cmidrule(rl){2-2} \cmidrule(rl){3-8} 

\multicolumn{1}{c}{ONNX + GPU} & \multicolumn{1}{c}{2.48}          & \multicolumn{1}{c}{0.57}   & \multicolumn{1}{c}{0.76}   & \multicolumn{1}{c}{0.92}   & \multicolumn{1}{c}{0.97}  & \multicolumn{1}{c}{1.21}   & 1.85   
\\ 

\multicolumn{1}{c}{ONNX + CPU} & \multicolumn{1}{c}{1.81}          & \multicolumn{1}{c}{0.72}  & \multicolumn{1}{c}{1.45}  & \multicolumn{1}{c}{2.48}  & \multicolumn{1}{c}{2.99}  & \multicolumn{1}{c}{3.32}  & 6.07
\\ \hline
\end{tabular}}}
\caption{Performance of the web-based viewer deployed by JavaScript library.}
\label{tab:1a}
\end{subtable}

\begin{subtable}{\columnwidth}\centering
{\resizebox{\columnwidth}{!}{
\begin{tabular}{cccccccc}
\hline
\multicolumn{8}{c}{ABC - $Lagrangian_{long}$ - 8.4 MB}              \\ 
\multicolumn{1}{c}{}  & \multicolumn{1}{c}{Model Loading (s)} & \multicolumn{1}{c}{\#100 (s)} & \multicolumn{1}{c}{\#200 (s)} & \multicolumn{1}{c}{\#300 (s)} & \multicolumn{1}{c}{\#400 (s)} & \multicolumn{1}{c}{\#500 (s)} & \#1000 (s) \\ 
\cmidrule(rl){2-2} \cmidrule(rl){3-8} 

\multicolumn{1}{c}{ONNX + GPU} & \multicolumn{1}{c}{1.15}          & \multicolumn{1}{c}{0.003} & \multicolumn{1}{c}{0.005} & \multicolumn{1}{c}{0.007} & \multicolumn{1}{c}{0.009} & \multicolumn{1}{c}{0.014} & 0.032  \\ 

\multicolumn{1}{c}{ONNX + CPU} & \multicolumn{1}{c}{0.072}         & \multicolumn{1}{c}{0.11}  & \multicolumn{1}{c}{0.22}  & \multicolumn{1}{c}{0.33}  & \multicolumn{1}{c}{0.41}  & \multicolumn{1}{c}{0.54}  & 0.95   \\ \hline

\multicolumn{8}{c}{Hurricane - $Lagrangian_{long}$ - 13.7 MB}        \\ 
\multicolumn{1}{c}{}  & \multicolumn{1}{c}{Model Loading (s)} & \multicolumn{1}{c}{\#100 (s)} & \multicolumn{1}{c}{\#200 (s)} & \multicolumn{1}{c}{\#300 (s)} & \multicolumn{1}{c}{\#400 (s)} & \multicolumn{1}{c}{\#500 (s)} & \#1000 (s) \\ 
\cmidrule(rl){2-2} \cmidrule(rl){3-8} 

\multicolumn{1}{c}{ONNX + GPU} & \multicolumn{1}{c}{1.18}          & \multicolumn{1}{c}{0.004} & \multicolumn{1}{c}{0.007} & \multicolumn{1}{c}{0.009} & \multicolumn{1}{c}{0.012} & \multicolumn{1}{c}{0.018} & 0.037  \\
\multicolumn{1}{c}{ONNX + CPU} & \multicolumn{1}{c}{0.11}          & \multicolumn{1}{c}{0.16}  & \multicolumn{1}{c}{0.26}  & \multicolumn{1}{c}{0.41}  & \multicolumn{1}{c}{0.54}  & \multicolumn{1}{c}{0.68}  & 1.29   \\ \hline
\end{tabular}}}
\caption{Performance of the OSPRay integration deployed by C++ library.}
\label{tab:1b}
\end{subtable}
\vspace{-1em}
\caption{The inference performance of our web-based viewer (\cref{tab:1a}) and OSPRay integration (\cref{tab:1b}) with increasing numbers of seeds (\#N represents the seed count) for the ABC and Hurricane datasets. We deploy the trained model using ONNX Runtime API. Performance is measured in seconds, and experiments are conducted on a workstation using a CPU ($20$ threads) or a GPU (CUDA). Our viewers enable interactive inference and visualization of new trajectories with the $Lagrangian_{long}$ method using the GPU. The C++ API is at least five times faster than the JavaScript API. Furthermore, although the GPU outperforms the CPU in inference, it is slower in loading the model.\vspace{-2em}}
\label{tab:1}
\end{table}

\cref{tab:1} presents the model loading and inference speeds of our web viewer and OSPRay integration using a CPU with $20$ threads and a GPU with CUDA on the workstation. 
With CUDA, our developed viewers enable full interactivity for up to $1,000$ new seeds in the inference pathlines. 
The motivation for deploying the trained neural network on a web browser is to enable users to perform post hoc exploration more easily, without requiring a high-performance computer. 
As shown in Tab.2 in the supplemental material, we observe similar performance results to those using the workstation (\cref{tab:1a}), indicating that the parallel algorithm used by ORT is not optimal. 
Therefore, further improvement and acceleration of web-based deployment is required. 
Moreover, although the web-based viewer with a CPU is not fully interactive, it is still much faster than the conventional interpolation approach. 


\subsection{Comparison with Interpolation Methods}
\label{sec:comparison}

To compute the trajectories of new start seeds, post hoc interpolation methods are applied after saving the basis Lagrangian flow maps through particle tracing. 
Conventional interpolation methods, including barycentric coordinate and Shepard's method, require the identification of the vicinity of the new seeds in the basis trajectories for computing a new particle trajectory. 
The methods for determining the neighborhoods rely on the structure of the basis flow maps. 
Delaunay triangulation can locate a cell in a structured or unstructured data source containing a Lagrangian representation. 

In our experiments, we compare the performance of our proposed approach to that of the conventional post hoc interpolation method (BC), which includes (1) loading the basis flow maps, (2) creating the triangulation structure, and (3) performing barycentric coordinate interpolation. 
In our implementation, we utilize the CGAL~\cite{fabri2009cgal} library to generate the triangulation structure and employ Threading Building Blocks~\cite{pheatt2008intel} (TBB) to parallelize all processes on CPUs. 
Our deep learning strategy (DL) consists of three steps: (1) loading the learned model, (2) generating the input using seed start locations and file cycles, and (3) inferring results using the training model. In our studies, we implement the network inference in C++ and utilize ORT~\footnote{https://onnxruntime.ai/docs/get-started/with-cpp.html} with CPU/GPU for the inference procedure of deep learning.
For evaluations, we use two structured datasets (Gerris ($Re=101.6$) and Hurricane) and the other two datasets (Heated Cylinder and Half Cylinder($Re=160$)) in both structured and unstructured formats. 
We compute the basis flow maps for structured datasets by placing seeds at each grid vertex (to add basis flow maps) and utilizing Sobol seeds to generate training data. 
In the case of unstructured data, we use sparse seeds at the center of each cell.

\begin{table}[h]
\centering
\resizebox{\columnwidth}{!}{
\begin{tabular}{ccccccc}
\hline
&  &  & \multicolumn{2}{c}{BC} & \multicolumn{2}{c}{DL} \\ 
\cmidrule(r){4-5} \cmidrule(l){6-7}
Datasets & \#FM & Resolution & Computation (hrs) & Storage (MB) & Training (hrs) & Storage (MB) \\
\cmidrule(r){1-3} \cmidrule(l){4-5} \cmidrule(l){6-7}
Gerris (Re 101.6) [S] & 100 & $512 \times 512$ &  0.072 & 683.8 & 0.93 &  10.3 \\
Hurricane [S] & 30 & $150 \times 150 \times 100$ & 0.15 & 1599 & 11.25 & 13.7 \\         
Heated Cylinder [U] & 100 & 49,610 &  0.03 & 101.3 & 0.25 & 41.59     \\
Heated Cylinder [S] & 100 & $150 \times 450$ & 0.03 & 137.4 & 3.45 & 41.59 \\

Half Cylinder (Re 160) [U] & 50 & 6,752  & 0.007 & 3.8 & 0.50 & 41.27 \\

Half Cylinder (Re 160) [S] & 50 & $80 \times 80 \times 80$ & 0.15 & 680.0 & 2.50 & 41.27 \\ 

\hline
\end{tabular}
}
\caption{\label{table:comparison} 
Comparing the computation time and storage requirements of the deep-learning-based approach (DL) vs. the conventional approach (BC). This table evaluates the time to compute basis flow maps and the duration of neural network training across various structured (S) and unstructured (U) datasets. It includes the number of flow maps ($\#FM$) and the resolution for each dataset. Additionally, it details the storage space needed for the basis flow map and the neural network. Even though our approach requires more time for neural network training, it consistently outperforms the conventional method in speed across all experiments (refer to \cref{table:comparison-loading} and \cref{fig:comparison_interpolation}) and reduces storage space requirements by two to 116 times compared to the conventional approach expect for the smallest unstructured Half Cylinder ($Re=160$).\vspace{-1em}}
\end{table}


\begin{table}[h]
\centering
\resizebox{\columnwidth}{!}{
\begin{tabular}{ccccc}
\hline
 & \multicolumn{2}{c}{BC} & \multicolumn{2}{c}{DL} \\ 
\cmidrule(r){2-3} \cmidrule(l){4-5}
Datasets & Loading (s) & Triangulation (s) & Loading w/ CPU (s) & Loading w/ GPU (s) \\
\cmidrule(r){1-1} \cmidrule(l){2-3} \cmidrule(l){4-5}
Gerris (Re 101.6) [S] & 2.6 & 6.67 & 0.06 & 1.63 \\
Hurricane [S] &  6.85 & 74.02 & 0.19 & 1.86 \\         
Heated Cylinder [U] & 0.53 & 1.27 & 0.19 & 1.76     \\
Heated Cylinder [S] & 1.55 & 1.74 & 0.19 & 1.76 \\

Half Cylinder (Re 160) [U] & 0.71 & 41.7 & 0.19 & 1.82 \\

Half Cylinder (Re 160) [S] & 4.13 & 29.47 & 0.19 & 1.82 \\ 

\hline
\end{tabular}
}
\caption{\label{table:comparison-loading} Computation time comparison of the deep-learning-based approach (DL) vs. the conventional approach (BC). This table illustrates the time required for various tasks in post hoc analysis: loading basis flow maps, triangulation for the conventional approach, and model loading for the deep learning approach, across different structured and unstructured datasets. Our results indicate that the DL approach is consistently faster than the BC approach in all experiments. Remarkably, for the Hurricane dataset, the processing time is reduced by up to $426$ when using CPUs and is approximately $44$ faster with a GPU. In the case of the smaller, unstructured Heated Cylinder dataset, network loading is nine times faster on a CPU, whereas the performance on a GPU is comparable. This result indicates that the DL approach effectively alleviates I/O constraints. The substantial speed advantage of the DL approach facilitates interactive flow field exploration, particularly in large 3D datasets.\vspace{-0.5em}}
\end{table}



\cref{table:comparison} illustrates the differences in computation time and storage requirements between the two approaches. The deep-learning-based method offers a reduction in storage needs by a factor of two to $116$ times when compared to the conventional approach, across all datasets. The only exception is the unstructured Half Cylinder dataset, which comprises a relatively small number of seeds, totaling $6,752$. 
Additionally, our approach is more storage efficient. For example, in the case of the Gerris flow dataset, storing the trained model results in a $68$-fold reduction in storage space. 
For the Hurricane dataset, implementing our model pruning strategy (\cref{sec:model_pruning}) enables a storage reduction by a factor of $116$. However, it is essential to note that for the smaller, unstructured Half Cylinder dataset, which contains only $6,752$  seeds, our approach necessitates increased storage compared to the other datasets.


In \cref{table:comparison-loading}, we compare the time required for the preparatory processes, including loading basis flow maps and performing triangulation in the BC interpolation approach, versus the model loading time in our proposed method. 
\cref{table:comparison-loading} demonstrates that our deep-learning-based strategy significantly reduces loading times. For the hurricane dataset, the processing is approximately $426$ times faster with a CPU and about $44$ times faster with a GPU. For the smaller, unstructured Heated Cylinder dataset, the network loading time is nine times faster with a CPU, and the performance is similar on a GPU, which indicates that our DL approach effectively mitigates I/O constraints. 
Especially for the structured datasets, our approach shows a significant speed advantage. The model loads in less than two seconds using a GPU and in under $0.2$ seconds with a CPU. This swift performance, in contrast to the longer duration required for loading basis flow maps in the BC approach, underscores the efficiency of our method during the initial setup phase.


\begin{figure}[!htb]
    \centering
    \vspace{-1em}
    \includegraphics[width=\columnwidth]{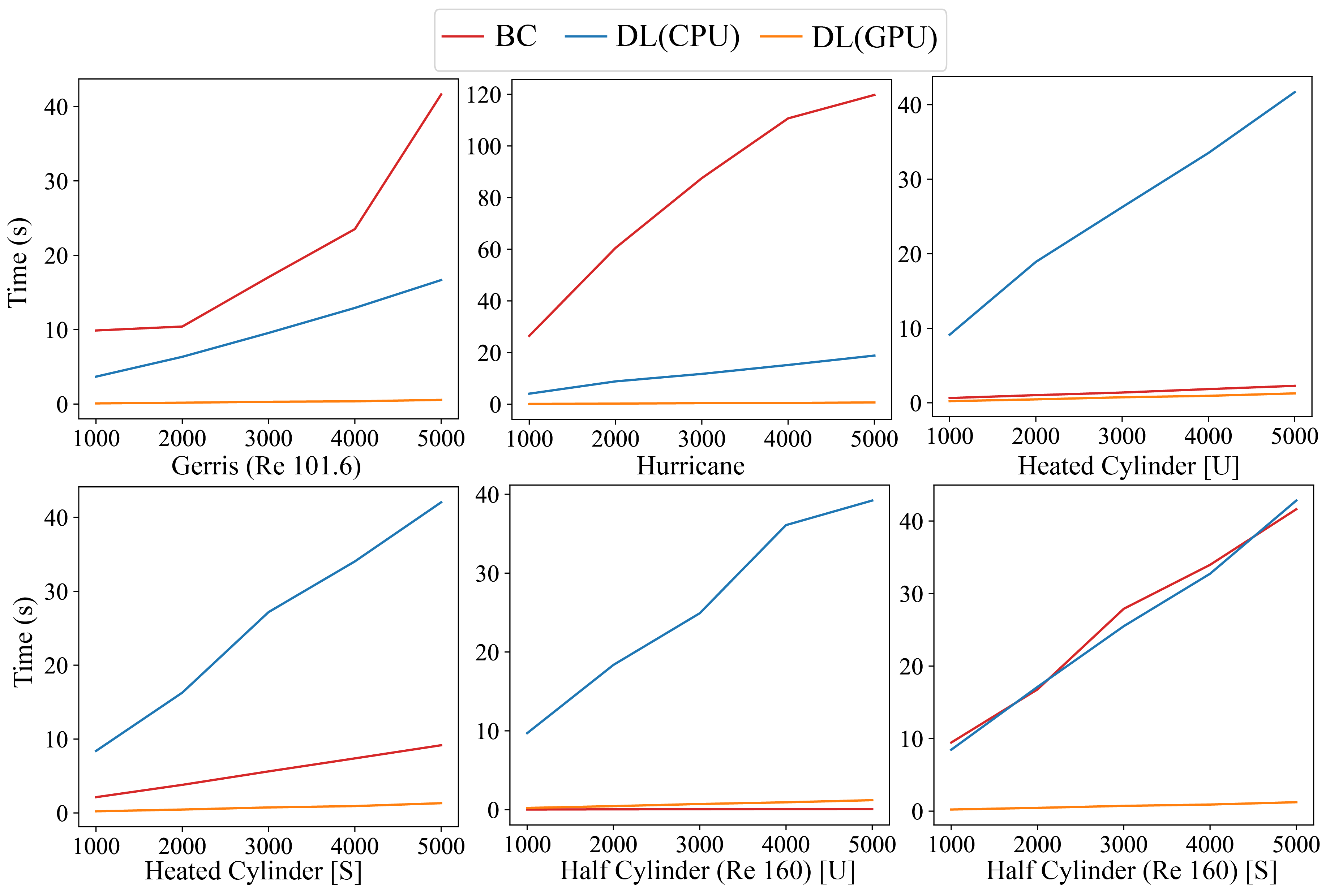}
    \caption{Comparison of post hoc interpolation times of deep-learning-based (DL) vs. conventional approach (BC). This chart compares the interpolation time for the conventional BC method and the inference time for the DL method across various structured and unstructured datasets, with an increasing number of seeds. Both CPU and GPU performance are evaluated for the DL method. Our findings show that the DL method, when utilizing a GPU, consistently outperforms the BC method in all tests involving structured datasets. It performs approximately $170$ times faster for the hurricane dataset, offering a substantial speed-up for interactive flow visualization and exploration in large-scale 3D datasets. When using a CPU, the DL method surpasses the BC method for high-resolution datasets like Gerris and Hurricane. It performs comparably to the BC method for the structured Half Cylinder dataset. Nonetheless, for all unstructured datasets, the DL method, whether employing GPU or CPU, is slower than the BC method.\vspace{-1em}}
    \label{fig:comparison_interpolation}
\end{figure}

\begin{figure}[!htb]
    \centering
    \includegraphics[width=\columnwidth]{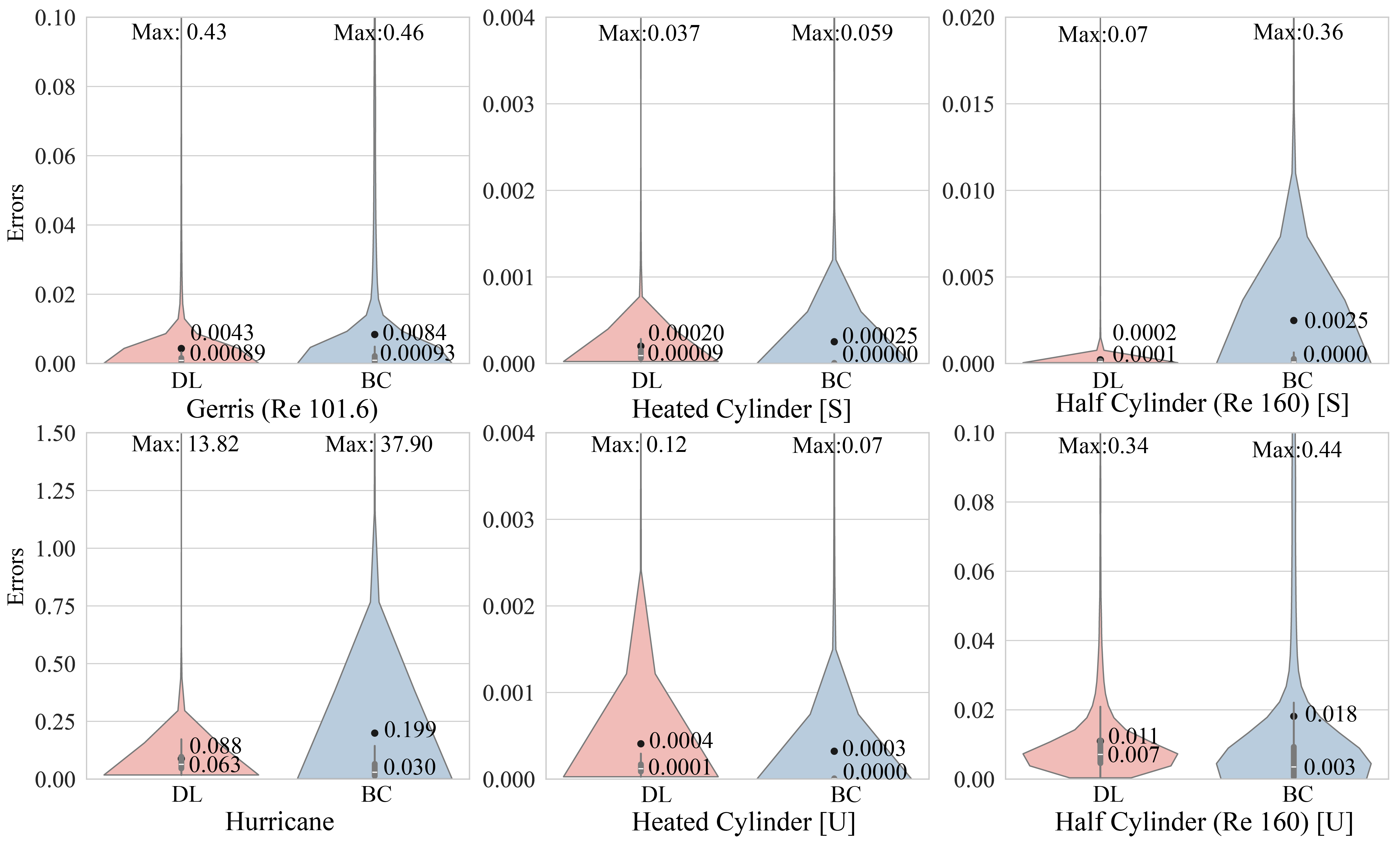}
    \vspace{-1em}
    \caption{The violin plot illustrates a comparison of error rates between our deep-learning-based method (DL) and the traditional barycentric coordinate interpolation (BC) method. 'S' represents structured data, whereas 'U' indicates unstructured data. The median is depicted by a white line on the gray bar, and the mean is shown as a black dot. The error range in each violin plot is displayed using error limits to clearly demonstrate the error distribution with the maximum error noted at the top of each plot. 
    In terms of median error, our method shows comparable results across all structured datasets. Notably, the BC method generates higher maximum errors compared to the DL method. For instance, the mean errors in our method are up to 12 times lower for the structured Half Cylinder dataset. However, our method tends to produce higher errors for unstructured datasets, especially in cases trained with sparse seeding. In order to develop a more accurate model for unstructured data, a mesh density-guided flow map sampling is necessary.} \vspace{-2em}
    \label{fig:comparison_errors}
\end{figure}

Our approach has shown promising results, but one limitation is the extensive training time required (refer to \cref{table:comparison}). 
For instance, the Hurricane dataset of $2.25$ million seeds and $30$ flow maps takes approximately $11$ hours to complete $100$ epochs during training. 
However, this is a common issue associated with deep learning and can be potentially mitigated by utilizing more powerful hardware or developing new training procedures in the future.

As depicted in \cref{fig:comparison_errors}, we evaluate the inference accuracy of our approach in comparison to the conventional approach using barycentric coordinates. Our method demonstrates comparable performance across all structured datasets when considering median errors. Significantly, the traditional method tends to yield higher maximum errors in comparison to our deep-learning-based method. For example, in the structured Half Cylinder dataset, the mean errors with our approach are up to 12 times lower.
However, our current model exhibits a larger error rate when dealing with unstructured data, particularly for 3D datasets such as the Half Cylinder, which employs sparse seeds for training. 
Therefore, to enhance the prediction accuracy of unstructured data, employing densely distributed seeds is crucial. Moreover, to overcome this limitation, we intend to explore an adaptive sampling strategy in our future work, which could aid in selecting important seeds for generating training data. More visual comparison results can be found in Fig.~2 and Fig.~3 of the supplemental material.


In summary, our proposed deep-learning-based method has the potential to substantially improve the prediction accuracy of structured datasets while also minimizing their storage requirements and expediting the interpolation procedure when compared to the conventional post hoc interpolation approach.



\subsection{Interactive Visualization Tool for Post Hoc Analysis}
\label{sec:viewer}

Benefiting from our model's minimal memory footprint and fast inference, we build a deep learning approach to accelerate the post hoc interpolation and visualization of Lagrangian flow maps, which is typically expensive due to I/O constraints and interpolation performance (\cref{sec:comparison}). 
To facilitate the exploration of flow maps without requiring a powerful computer, we develop a web-based viewer that utilizes our neural network as the backend to accelerate the interpolation and visualization process. 
The viewer is implemented in JavaScript and is compatible with multiple platforms and popular web browsers, such as Safari, Chrome, and Firefox. 
The user interface of the web-based viewer includes control panels for model loading, seed placement, seed box configuration, scalars configuration, transfer function editing, and tracing. 
It also provides a primary viewer for presenting the visualization of pathlines, surfaces, and volumes, allowing users to engage with seed placement and visualization outcomes (\cref{fig:abc_web_vis}). 
The details of the implementation of the web viewer, as outlined in Sec.~4 of the supplemental material, along with the visualization results presented in Fig. 6 and 7, are further elaborated upon in the supplemental materials.   

In addition to the web-based viewer, we have also integrated our neural network with the OSPRay~\cite{wald2016ospray} rendering engine to enable C++ implementation (see supplemental material). 
Our integration with the web-based viewer and OSPRay rendering engine is independent of the model architecture, allowing users to integrate their models easily by simply replacing the trace function. 
We highlight the visualization result of \textbf{ScalarFlow} in \cref{fig:ospray}.  

All of our source code is available on GitHub.\footnote{\url{https://github.com/MengjiaoH/FlowMap_Web_Viewer}} \footnote{\url{https://github.com/MengjiaoH/FlowMap_OSPRay_Viewer}}

\begin{figure}[!htb]
    \centering
    \includegraphics[width=\columnwidth]{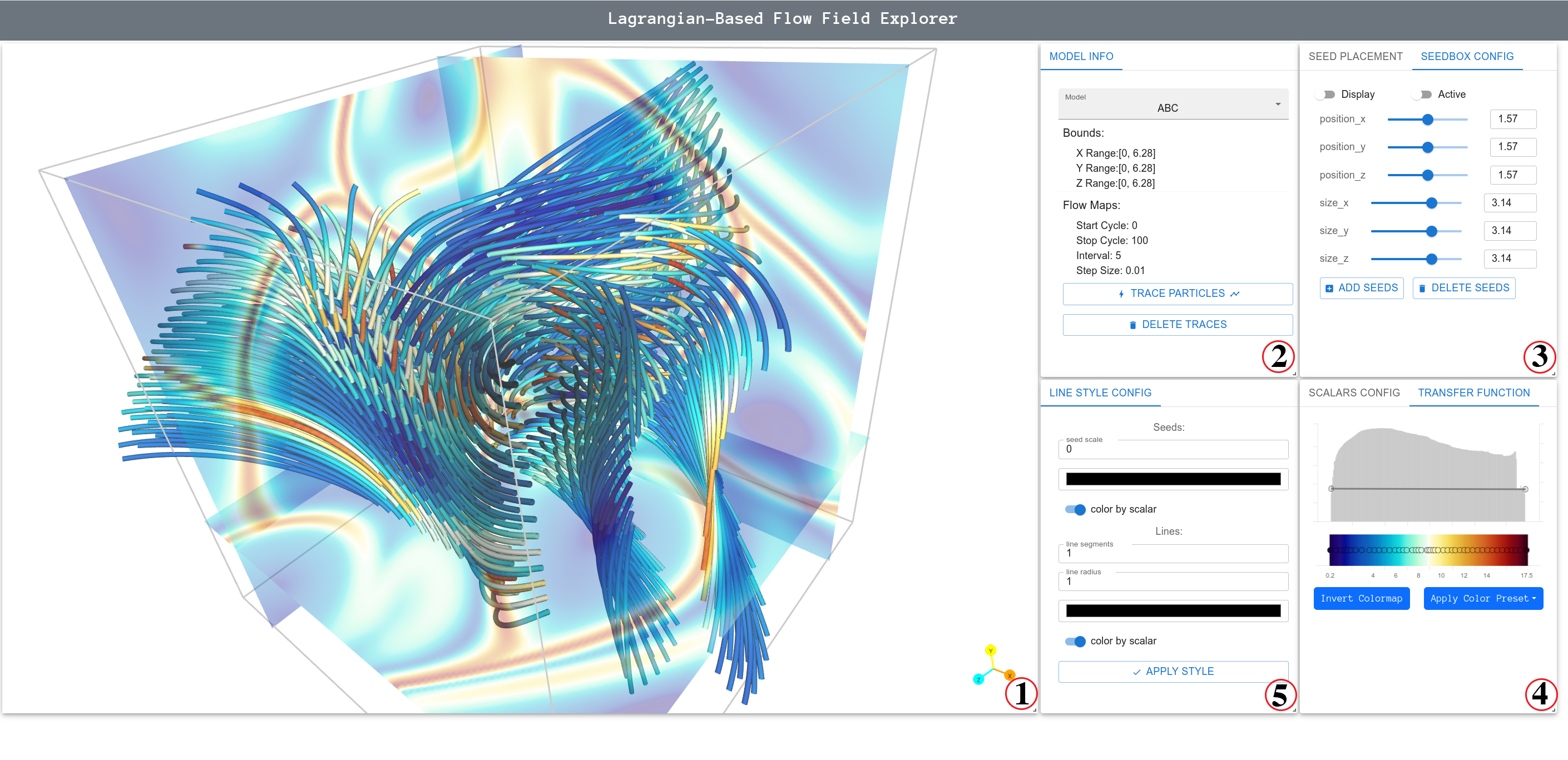}
    \vspace{-2em}
    \caption{Illustration of our web-based viewer for visualizing inferred pathlines using our pre-trained model in the \textbf{ABC} dataset. The interface includes panels for (1) main display, (2) model loading, data information and particle tracing, (3) seed box configuration, (4) transfer function for scalar field data visualization, and (5) seed and line style configuration.    \vspace{-2em}}

    \label{fig:abc_web_vis}
\end{figure}


\begin{figure}[!htb]
    \centering
    \includegraphics[width=0.8\linewidth, keepaspectratio]{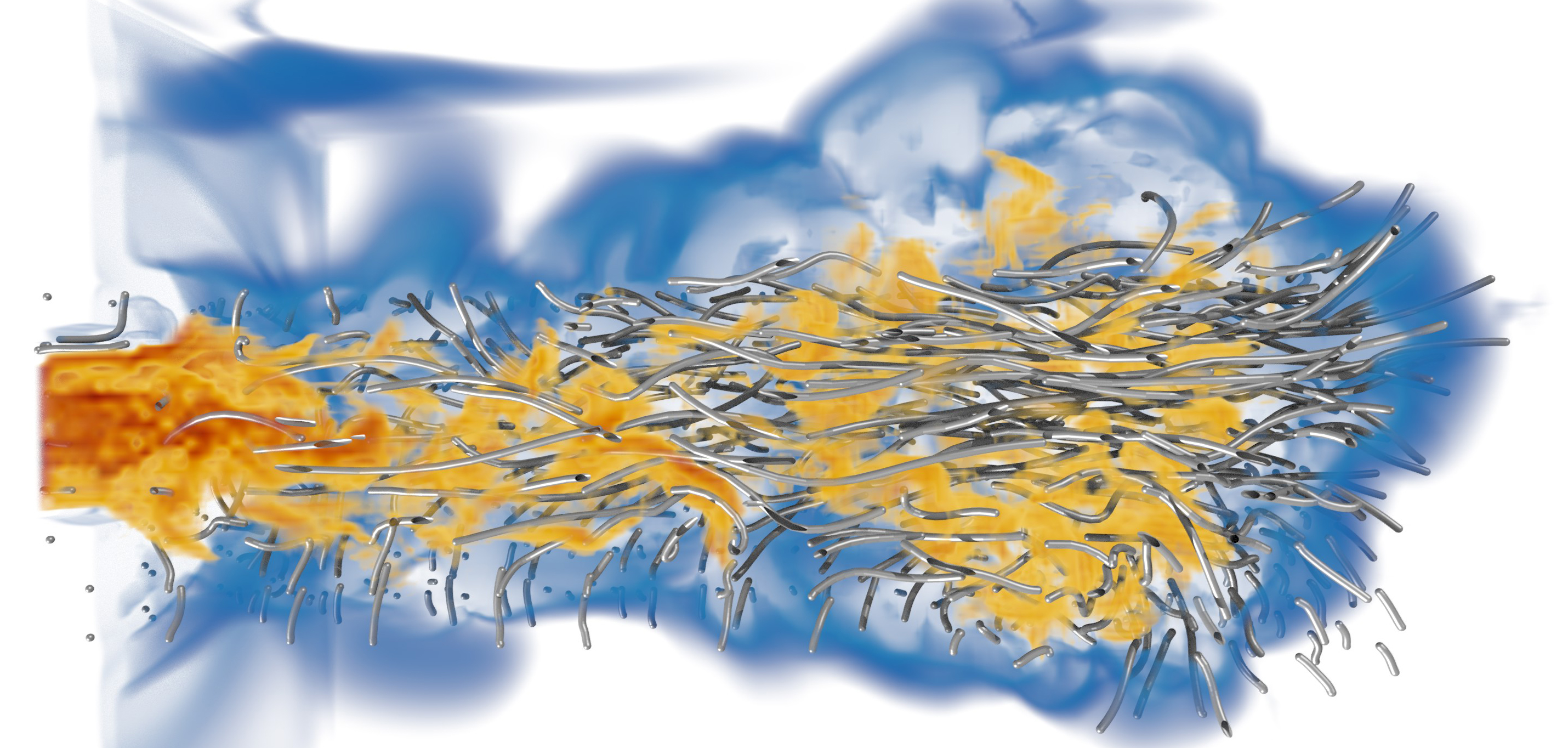}
    \caption{Multiworkflow visualization of the \textbf{ScalarFlow} dataset using our OSPRay-based viewer, which integrates our neural network with the OSPRay renderer. The visualization displays the FTLE as a volume and the pathlines inferred using our neural network. A clipping plane is aligned along the y-axis. The model is trained with the $Lagrangian_{hybrid}$ method. Each pathline encompasses $15$ time steps, ranging from time step $135$ to time step $150$.\vspace{-2em}}
    \label{fig:ospray}
\end{figure}

\section{Conclusion and Future Work}
\label{sec:conclusion}

In this paper, we conduct a comprehensive evaluation of a deep learning approach that uses Lagrangian representations to accelerate the post hoc interpolation and visualization of time-varying flow fields.

Our results help identify best practices for using MLP-based models to reconstruct Lagrangian flow maps. Our findings include:
\begin{itemize}[noitemsep]
    \item Shallow neural networks generally perform better than deep neural networks in reconstructing  Lagrangian flow maps; 
    \item The reconstruction errors tend to increase with increased turbulence in flow behaviors. A larger hidden layer (with a higher latent dimension) is more effective than a smaller one for capturing complex flow dynamics; 
    \item The Sinusoidal activation function demonstrates superior performance compared to the ReLU activation function; 
    \item Employing the $Lagrangian_{hybrid}$ method for generating training data effectively minimizes error propagation while preserving domain coverage, in contrast to the $Lagrangian_{long}$; 
    \item Model pruning is essential for reducing model size and enhancing the efficiency of the inference process.
\end{itemize}

By comparing the deep learning approach to the conventional method based on Delaunay triangulation and barycentric coordinate interpolation, we demonstrate that our approach is at least three times faster on small datasets and over 400 times faster on large 3D datasets. 
Our method improves interpolation accuracy by threefold on average for structured datasets. 
In terms of storage space, our method reduces memory usage by $46$ times for the Hurricane dataset without model pruning, by saving $30$ basis flow maps with 1.6 GB storage. With the application of model pruning techniques, memory usage is further reduced, achieving a 116-fold reduction.

By leveraging the rapid inference and small memory footprint of our MLP model, we provide a web-based viewer to offer an easy way to visualize and explore the post hoc interpolation process. 
Accessible from any computer regardless of processing capacity, the web-based viewer supports multiple platforms and web browsers. It also provides various seeding strategies and supports both volume and slice representations of other scalar fields, enabling users to select regions of interest and explore them interactively.

When using CUDA-supported devices, our approach can generate pathlines interactively. Inferring $1,000$ pathlines with $30$ flow maps takes only one second. 
However, on a low-end device with a single CPU, the inference speed is slower, requiring three seconds to infer $1,000$ pathlines with $30$ flow maps. Despite the slower performance on low-end devices, interactive pathline generation remains significantly faster than the conventional approach. 
Whereas the current parallel scheme for deploying the neural network on the website using a CPU is suboptimal, we are optimistic that performance can be improved with a more efficient parallel API in the future. 

Additionally, we have integrated our neural network into the OSPRay rendering engine to support C++ developers and enable fast, high-fidelity rendering performance. Our integration is general and can be applied to other model architectures with similar tasks. 
We have made source code available on GitHub, allowing users to easily integrate their neural networks by replacing the trace function.

In addition, our study demonstrates that the proposed neural network, featuring a sinusoidal activation function, significantly improves inference accuracy compared to a prior study~\cite{han2021exploratory}. We also investigate the impact of different model architectures on various 2D and 3D datasets, and assess our model's ability to handle datasets with increasing flow complexity. Moreover, we evaluate a $Lagrangian_{hybrid}$ training data structure that conceptually combines the $Lagrangian_{long}$ and $Lagrangian_{short}$ methods utilized in~\cite{han2021exploratory}, ensuring domain convergence while minimizing error propagation.

Even though our approach has shown promising results, certain limitations exist.
We observe larger errors in unstructured datasets with sparse seeds. To address this limitation, we plan to investigate adaptive seeding approaches that place seeds based on flow features and mesh refinement instead of uniform seeding. 
Additionally, training the model for large 3D datasets is computationally expensive. We plan to explore ways to reduce training time while maintaining accuracy, such as utilizing transfer learning. Furthermore, other techniques for encoding high-frequency signals, such as positional encoding, could be explored alongside sinusoidal activation functions to further enhance accuracy.
Lastly, our model is less effective at inferring long trajectories, especially in complex flow behavior. Improving the model's capability to handle large-scale datasets is also an area of future research.

\acknowledgments{
\vspace{-0.5em}
This work was partially supported by the Intel OneAPI CoE, the Intel Graphics and Visualization Institutes of XeLLENCE, and the DOE Ab-initio Visualization for Innovative Science (AIVIS) grant 2428225. Additional support comes from NSF SHF award 2221812 and DOE DE-SC0023157.
\vspace{-1em}
}

\bibliographystyle{abbrv}
\bibliography{flow-vis-main.bib}

\begin{thebibliography}{10}

\bibitem{agranovsky2014improved}
A.~Agranovsky, D.~Camp, C.~Garth, E.~W. Bethel, K.~I. Joy, and H.~Childs.
\newblock {Improved} {Post} {Hoc} {Flow} {Analysis} {Via} {Lagrangian}
  {Representations}.
\newblock In {\em 2014 IEEE 4th Symposium on Large Data Analysis and
  Visualization (LDAV)}, pages 67--75, 2014.

\bibitem{agranovsky2015multi}
A.~Agranovsky, H.~Obermaier, C.~Garth, and K.~I. Joy.
\newblock A {M}ulti-{R}esolution {I}nterpolation {S}cheme for {P}athline
  {B}ased {L}agrangian {F}low {R}epresentations.
\newblock In {\em Visualization and Data Analysis 2015}, volume 9397, page
  93970K, 2015.

\bibitem{ali2010visualization}
A.~S. Ali, A.~S. Hussien, M.~F. Tolba, and A.~H. Youssef.
\newblock {V}isualization of {L}arge {T}ime-{V}arying {V}ector {D}ata.
\newblock In {\em 2010 3rd International Conference on Computer Science and
  Information Technology}, volume~4, pages 210--215. IEEE, 2010.

\bibitem{an2021stsrnet}
Y.~An, H.-W. Shen, G.~Shan, G.~Li, and J.~Liu.
\newblock {S}{T}{S}{R}{N}et: {D}eep {J}oint {S}pace-{T}ime {S}uper-{R}esolution
  for {V}ector {F}ield {V}isualization.
\newblock {\em IEEE Computer Graphics and Applications}, 41(6):122--132, 2021.

\bibitem{BaezaRojo19SciVisa}
I.~Baeza~Rojo and T.~G{\"u}nther.
\newblock {V}ector {F}ield {T}opology of {T}ime-{D}ependent {F}lows in a
  {S}teady {R}eference {F}rame.
\newblock {\em IEEE Transactions on Visualization and Computer Graphics (Proc.
  IEEE Scientific Visualization)}, 2019.

\bibitem{bai2019streampath}
X.~Bai, C.~Wang, and C.~Li.
\newblock {A} {Streampath}-{Based} {RCNN} {Approach} to {Ocean} {Eddy}
  {Detection}.
\newblock {\em IEEE Access}, 7:106336--106345, 2019.

\bibitem{beck2020neural}
A.~D. Beck, J.~Zeifang, A.~Schwarz, and D.~G. Flad.
\newblock A {N}eural {N}etwork based {S}hock {D}etection and {L}ocalization
  {A}pproach for {D}iscontinuous {G}alerkin {M}ethods.
\newblock {\em Journal of Computational Physics}, 423:109824, 2020.

\bibitem{brunton2020machine}
S.~L. Brunton, B.~R. Noack, and P.~Koumoutsakos.
\newblock {M}achine {L}earning for {F}luid {M}echanics.
\newblock {\em Annual Review of Fluid Mechanics}, 52:477--508, 2020.

\bibitem{bujack2015lagrangian}
R.~Bujack and K.~I. Joy.
\newblock {L}agrangian {R}epresentations of {F}low {F}ields with {P}arameter
  {C}urves.
\newblock In {\em 2015 IEEE 5th Symposium on Large Data Analysis and
  Visualization (LDAV)}, pages 41--48. IEEE, 2015.

\bibitem{chandler2014interpolation}
J.~Chandler, H.~Obermaier, and K.~I. Joy.
\newblock Interpolation-{B}ased {P}athline {T}racing in {P}article-{B}ased
  {F}low {V}isualization.
\newblock {\em IEEE transactions on visualization and computer graphics},
  21(1):68--80, 2014.

\bibitem{da2004lagrangian}
M.~V. Da~Costa and B.~Blanke.
\newblock {Lagrangian methods for flow climatologies and trajectory error
  assessment}.
\newblock {\em Ocean Modelling}, 6(3-4):335--358, 2004.

\bibitem{deng2019cnn}
L.~Deng, Y.~Wang, Y.~Liu, F.~Wang, S.~Li, and J.~Liu.
\newblock {A} {CNN}-based {Vortex} {Identification} {Method}.
\newblock {\em Journal of Visualization}, 22(1):65--78, 2019.

\bibitem{dong2015image}
C.~Dong, C.~C. Loy, K.~He, and X.~Tang.
\newblock {Image} {Super}-{Resolution} {Using} {Deep} {Convolutional}
  {Networks}.
\newblock {\em IEEE transactions on pattern analysis and machine intelligence},
  38(2):295--307, 2015.

\bibitem{duo2019oceanic}
Z.~Duo, W.~Wang, and H.~Wang.
\newblock {Oceanic} {Mesoscale} {Eddy} {Detection} {Method} {Based} on {Deep}
  {Learning}.
\newblock {\em Remote Sensing}, 11(16):1921, 2019.

\bibitem{eckert2019scalarflow}
M.-L. Eckert, K.~Um, and N.~Thuerey.
\newblock {S}calar{F}low: {A} {L}arge-{S}cale {V}olumetric {D}ata {S}et of
  {R}eal-world {S}calar {T}ransport {F}lows for {C}omputer {A}nimation and
  {M}achine {L}earning.
\newblock {\em ACM Transactions on Graphics (TOG)}, 38(6):1--16, 2019.

\bibitem{elesedy2020lottery}
B.~Elesedy, V.~Kanade, and Y.~W. Teh.
\newblock {L}ottery {T}ickets in {L}inear {M}odels: {A}n {A}nalysis of
  {I}terative {M}agnitude {P}runing.
\newblock {\em arXiv preprint arXiv:2007.08243}, 2020.

\bibitem{fabri2009cgal}
A.~Fabri and S.~Pion.
\newblock {C}{G}{A}{L}: {T}he {C}omputational {G}eometry {A}lgorithms
  {L}ibrary.
\newblock In {\em Proceedings of the 17th ACM SIGSPATIAL international
  conference on advances in geographic information systems}, pages 538--539,
  2009.

\bibitem{fang2023depgraph}
G.~Fang, X.~Ma, M.~Song, M.~B. Mi, and X.~Wang.
\newblock {D}ep{G}raph: {T}owards {A}ny {S}tructural {P}runing.
\newblock {\em arXiv preprint arXiv:2301.12900}, 2023.

\bibitem{frankle2018lottery}
J.~Frankle and M.~Carbin.
\newblock {T}he {L}ottery {T}icket {H}ypothesis: {F}inding {S}parse,
  {T}rainable {N}eural {N}etworks.
\newblock {\em arXiv preprint arXiv:1803.03635}, 2018.

\bibitem{franz2018ocean}
K.~Franz, R.~Roscher, A.~Milioto, S.~Wenzel, and J.~Kusche.
\newblock Ocean eddy identification and tracking using neural networks.
\newblock In {\em IGARSS 2018-2018 IEEE International Geoscience and Remote
  Sensing Symposium}, pages 6887--6890. IEEE, 2018.

\bibitem{froyland2018robust}
G.~Froyland and O.~Junge.
\newblock {Robust} {FEM}-{Based} {Extraction} of {Finite}-{Time} {Coherent}
  {Sets} {Using} {Scattered}, {Sparse}, and {Incomplete} {Trajectories}.
\newblock {\em SIAM Journal on Applied Dynamical Systems}, 17(2):1891--1924,
  2018.

\bibitem{froyland2015rough}
G.~Froyland and K.~Padberg-Gehle.
\newblock A rough-and-ready cluster-based approach for extracting finite-time
  coherent sets from sparse and incomplete trajectory data.
\newblock {\em Chaos: An Interdisciplinary Journal of Nonlinear Science},
  25(8):087406, 2015.

\bibitem{gao2021super}
H.~Gao, L.~Sun, and J.-X. Wang.
\newblock {S}uper-resolution and denoising of fluid flow using physics-informed
  convolutional neural networks without high-resolution labels.
\newblock {\em Physics of Fluids}, 33(7):073603, 2021.

\bibitem{Guenther17}
T.~G{\"u}nther, M.~Gross, and H.~Theisel.
\newblock {G}eneric {O}bjective {V}ortices for {F}low {V}isualization.
\newblock {\em ACM Transactions on Graphics (Proc. SIGGRAPH)},
  36(4):141:1--141:11, 2017.

\bibitem{guo2020ssr}
L.~Guo, S.~Ye, J.~Han, H.~Zheng, H.~Gao, D.~Z. Chen, J.-X. Wang, and C.~Wang.
\newblock {S}{S}{R}-{V}{F}{D}: {Spatial} {Super}-{Resolution} for {Vector}
  {Field} {Data} {Analysis} and {Visualization}.
\newblock In {\em 2020 IEEE Pacific Visualization Symposium (PacificVis)},
  pages 71--80. IEEE Computer Society, 2020.

\bibitem{hadjighasem2017critical}
A.~Hadjighasem, M.~Farazmand, D.~Blazevski, G.~Froyland, and G.~Haller.
\newblock A {Critical} {Comparison} of {Lagrangian} {Methods} for {Coherent}
  {Structure} {Detection}.
\newblock {\em Chaos: An Interdisciplinary Journal of Nonlinear Science},
  27(5):053104, 2017.

\bibitem{han2018flownet}
J.~Han, J.~Tao, and C.~Wang.
\newblock {F}low{N}et: {A} {D}eep {L}earning {F}ramework for {C}lustering and
  {S}election of {S}treamlines and {S}tream {S}urfaces.
\newblock {\em IEEE transactions on visualization and computer graphics},
  26(4):1732--1744, 2018.

\bibitem{han2019flow}
J.~Han, J.~Tao, H.~Zheng, H.~Guo, D.~Z. Chen, and C.~Wang.
\newblock {Flow} {F}ield {R}eduction {V}ia {R}econstructing {V}ector {D}ata
  {F}rom 3-{D} {S}treamlines {U}sing {D}eep {L}earning.
\newblock {\em IEEE computer graphics and applications}, 39(4):54--67, 2019.

\bibitem{han2022tsr}
J.~Han and C.~Wang.
\newblock {T}{S}{R}-{V}{F}{D}: {G}enerating {T}emporal {S}uper-{R}esolution for
  {U}nsteady {V}ector {F}ield {D}ata.
\newblock {\em Computers \& Graphics}, 103:168--179, 2022.

\bibitem{han2021exploratory}
M.~Han, S.~Sane, and C.~R. Johnson.
\newblock Exploratory {Lagrangian}-{B}ased {P}article {T}racing {U}sing {D}eep
  {L}earning.
\newblock {\em Journal of Flow Visualization and Image Processing}, 2022.

\bibitem{helgeland2006high}
A.~Helgeland and T.~Elboth.
\newblock {H}igh-{Q}uality and {I}nteractive {A}nimations of 3{D}
  {T}ime-{V}arying {V}ector {F}ields.
\newblock {\em IEEE Transactions on Visualization and Computer Graphics},
  12(6):1535--1546, 2006.

\bibitem{hohlein2020comparative}
K.~H{\"o}hlein, M.~Kern, T.~Hewson, and R.~Westermann.
\newblock A comparative study of convolutional neural network models for wind
  field downscaling.
\newblock {\em Meteorological Applications}, 27(6):e1961, 2020.

\bibitem{hong2018access}
F.~Hong, J.~Zhang, and X.~Yuan.
\newblock {Access} {Pattern} {Learning} with {Long} {Short}-{Term} {Memory} for
  {Parallel} {Particle} {Tracing}.
\newblock In {\em 2018 IEEE Pacific Visualization Symposium (PacificVis)},
  pages 76--85. IEEE, 2018.

\bibitem{Jakob2020}
J.~Jakob, M.~Gross, and T.~G{\"u}nther.
\newblock {A} {Fluid} {Flow} {Data} {Set} for {Machine} {Learning} and its
  {Application} to {Neural} {Flow} {Map} {Interpolation}.
\newblock {\em IEEE Transactions on Visualization and Computer Graphics},
  27(2):1279--1289, 2020.

\bibitem{kashir2021application}
B.~Kashir, M.~Ragone, A.~Ramasubramanian, V.~Yurkiv, and F.~Mashayek.
\newblock Application of fully convolutional neural networks for feature
  extraction in fluid flow.
\newblock {\em Journal of Visualization}, 24(4):771--785, 2021.

\bibitem{kim2019robust}
B.~Kim and T.~G{\"u}nther.
\newblock {Robust} {Reference} {Frame} {Extraction} from {Unsteady} 2{D}
  {Vector} {Fields} with {Convolutional} {Neural} {Networks}.
\newblock In {\em Computer Graphics Forum}, volume~38, pages 285--295. Wiley
  Online Library, 2019.

\bibitem{lee2021deep}
J.-Y. Lee and J.~Park.
\newblock {Deep} {Regression} {Network}-{Assisted} {Efficient} {Streamline}
  {Generation} {Method}.
\newblock {\em IEEE Access}, 9:111704--111717, 2021.

\bibitem{lguensat2018eddynet}
R.~Lguensat, M.~Sun, R.~Fablet, P.~Tandeo, E.~Mason, and G.~Chen.
\newblock {EddyNet}: {A} {Deep} {Neural} {Network} for {Pixel}-{Wise}
  {Classification} of {Oceanic} {Eddies}.
\newblock In {\em IGARSS 2018-2018 IEEE International Geoscience and Remote
  Sensing Symposium}, pages 1764--1767. IEEE, 2018.

\bibitem{li2015extracting}
Y.~Li, C.~Wang, and C.-K. Shene.
\newblock {Extracting} {Flow} {Features} via {Supervised} {Streamline}
  {Segmentation}.
\newblock {\em Computers \& Graphics}, 52:79--92, 2015.

\bibitem{liu2022deep}
C.~Liu, R.~Jiang, D.~Wei, C.~Yang, Y.~Li, F.~Wang, and X.~Yuan.
\newblock {D}eep {L}earning {A}pproaches in {F}low {V}isualization.
\newblock {\em Advances in Aerodynamics}, 4(1):1--14, 2022.

\bibitem{liu2019cnn}
Y.~Liu, Y.~Lu, Y.~Wang, D.~Sun, L.~Deng, F.~Wang, and Y.~Lei.
\newblock A {C}{N}{N}-based shock detection method in flow visualization.
\newblock {\em Computers \& Fluids}, 184:1--9, 2019.

\bibitem{pheatt2008intel}
C.~Pheatt.
\newblock {I}ntel{\textregistered} {T}hreading {B}uilding {B}locks.
\newblock {\em Journal of Computing Sciences in Colleges}, 23(4):298--298,
  2008.

\bibitem{gerrisflowsolver}
S.~Popinet.
\newblock {F}ree {C}omputational {F}luid {D}ynamics.
\newblock {\em ClusterWorld}, 2(6), 2004.

\bibitem{Qin2014}
X.~Qin, E.~van Sebille, and A.~S. Gupta.
\newblock Quantification of errors induced by temporal resolution on
  {L}agrangian particles in an eddy-resolving model.
\newblock {\em Ocean Modelling}, 76:20--30, 2014.

\bibitem{rapp2019void}
T.~Rapp, C.~Peters, and C.~Dachsbacher.
\newblock {V}oid-and-{C}luster {S}ampling of {L}arge {S}cattered {D}ata and
  {T}rajectories.
\newblock {\em IEEE transactions on visualization and computer graphics},
  26(1):780--789, 2019.

\bibitem{rockwood2019practical}
M.~P. Rockwood, T.~Loiselle, and M.~A. Green.
\newblock Practical concerns of implementing a finite-time lyapunov exponent
  analysis with under-resolved data.
\newblock {\em Experiments in Fluids}, 60(4):1--16, 2019.

\bibitem{sahoo2021integration}
S.~Sahoo and M.~Berger.
\newblock {Integration}-{Aware} {Vector} {Field} {Super} {Resolution}.
\newblock 2021.

\bibitem{sahoo2022neural}
S.~Sahoo, Y.~Lu, and M.~Berger.
\newblock Neural {F}low {M}ap {R}econstruction.
\newblock In {\em Computer Graphics Forum}, volume~41, pages 391--402. Wiley
  Online Library, 2022.

\bibitem{sane2018revisiting}
S.~Sane, R.~Bujack, and H.~Childs.
\newblock Revisiting the {E}valuation of {I}n {S}itu {L}agrangian {A}nalysis.
\newblock In {\em EGPGV@ EuroVis}, pages 63--67, 2018.

\bibitem{sane2020survey}
S.~Sane, R.~Bujack, C.~Garth, and H.~Childs.
\newblock A {S}urvey of {S}eed {P}lacement and {S}treamline {S}election
  {T}echniques.
\newblock In {\em Computer Graphics Forum}, volume~39, pages 785--809. Wiley
  Online Library, 2020.

\bibitem{sane2022exploratory}
S.~Sane and H.~Childs.
\newblock {E}xploratory {T}ime-{D}ependent {F}low {V}isualization via {I}n
  {S}itu {E}xtracted {L}agrangian {R}ßepresentations.
\newblock In {\em In Situ Visualization for Computational Science}, pages
  91--109. Springer, 2022.

\bibitem{sane2019interpolation}
S.~Sane, H.~Childs, and R.~Bujack.
\newblock {An {I}nterpolation {S}cheme for {V}{D}{V}{P} {L}agrangian {B}asis
  {F}lows}.
\newblock In {\em Eurographics Symposium on Parallel Graphics and
  Visualization}, pages 109--119, 2019.

\bibitem{sane2021investigating}
S.~Sane, C.~R. Johnson, and H.~Childs.
\newblock {I}nvestigating {I}n {S}itu {R}eduction via {L}agrangian
  {R}epresentations for {C}osmology and {S}eismology {A}pplications.
\newblock In {\em International Conference on Computational Science}, pages
  436--450. Springer, 2021.

\bibitem{schlueter2017coherent}
K.~L. Schlueter-Kuck and J.~O. Dabiri.
\newblock Coherent structure colouring: identification of coherent structures
  from sparse data using graph theory.
\newblock {\em Journal of Fluid Mechanics}, 811:468--486, 2017.

\bibitem{shi2016real}
W.~Shi, J.~Caballero, F.~Husz{\'a}r, J.~Totz, A.~P. Aitken, R.~Bishop,
  D.~Rueckert, and Z.~Wang.
\newblock {Real}-{Time} {Single} {Image} and {Video} {Super}-{Resolution}
  {Using} an {Efficient} {Sub}-{Pixel} {Convolutional} {Neural} {Network}.
\newblock In {\em Proceedings of the IEEE conference on computer vision and
  pattern recognition}, pages 1874--1883, 2016.

\bibitem{sitzmann2020implicit}
V.~Sitzmann, J.~Martel, A.~Bergman, D.~Lindell, and G.~Wetzstein.
\newblock {I}mplicit {N}eural {R}epresentations with {P}eriodic {A}ctivation
  {F}unctions.
\newblock {\em Advances in Neural Information Processing Systems}, 33, 2020.

\bibitem{strofer2018data}
C.~M. Str{\"o}fer, J.~Wu, H.~Xiao, and E.~Paterson.
\newblock {Data}-{Driven}, {Physics}-{Based} {Feature} {Extraction} from
  {Fluid} {Flow} {Fields} {Using} {Convolutional} {Neural} {Networks}.
\newblock {\em Communications in Computational Physics}, 25(3):625--650, 2018.

\bibitem{tatarenkova2020edge}
D.~Tatarenkova.
\newblock Edge {D}etection and {M}achine {L}earning {A}pproach to {I}dentify
  {F}low {S}tructures on {S}chlieren and {S}hadowgraph {I}mages.
\newblock 2020.

\bibitem{wald2016ospray}
I.~Wald, G.~P. Johnson, J.~Amstutz, C.~Brownlee, A.~Knoll, J.~Jeffers,
  J.~G{\"u}nther, and P.~Navr{\'a}til.
\newblock {O}{S}{P}{R}ay-{A} {C}{P}{U} {R}ay {T}racing {F}ramework for
  {S}cientific {V}isualization.
\newblock {\em IEEE transactions on visualization and computer graphics},
  23(1):931--940, 2016.

\bibitem{wang2022dl4scivis}
C.~Wang and J.~Han.
\newblock {D}{L}4{S}ci{V}is: {A} {S}tate-of-the-{A}rt {S}urvey on {D}eep
  {L}earning for {S}cientific {V}isualization.
\newblock {\em IEEE Transactions on Visualization and Computer Graphics}, 2022.

\bibitem{wang2021rapid}
Y.~Wang, L.~Deng, Z.~Yang, D.~Zhao, and F.~Wang.
\newblock {A} {Rapid} {Vortex} {Identification} {Method} {Using} {Fully}
  {Convolutional} {Segmentation} {Network}.
\newblock {\em The Visual Computer}, 37(2):261--273, 2021.

\bibitem{yi2018cnn}
T.~B.~L. Yi.
\newblock {CNN}-based {Flow} {Field} {Feature} {Visualization} {Method}.
\newblock {\em International Journal of Performability Engineering}, 14(3):434,
  2018.

\end{thebibliography}

\end{document}